\documentclass[cernpreprint, orcidlogo, atlasdraft=false, texlive=2023, USenglish, texmf, orcidlogo]{atlasdoc}
 
\usepackage[subcaption]{atlaspackage}
 
\usepackage{physics}
 
\usepackage{atlasbiblatex}
 
\usepackage[unit=false, hepparticle=true, hepprocess=true, jetetmiss=true, misc=true]{atlasphysics}
 
\graphicspath{{logos/}{figs/}}

\newif\ifPrintNotes
\PrintNotesfalse  
 
\usepackage{ANA-TOPQ-2019-06-PAPER-defs}

\AtlasTitle{Search for flavor-changing neutral-current couplings
between the top quark and the $Z$ boson with proton--proton collisions at
$\sqrt{s} = 13\,\text{TeV}$ with the ATLAS detector}

\PreprintIdNumber{CERN-EP-2022-044}
 
\AtlasJournalRef{Phys.\ Rev.\ D 108 (2023) 032019}
\AtlasDOI{10.1103/PhysRevD.108.032019}
 
\AtlasAbstract{
A search for flavor-changing neutral-current couplings between a
top quark, an up or charm quark and a $Z$ boson is presented,
using proton--proton collision data at $\sqrt{s} = 13\,\text{TeV}$
collected by the ATLAS detector at the Large Hadron Collider.
The analyzed dataset corresponds to an integrated luminosity of
\SI{139}{\per\fb}.
The search targets both single-top-quark events
produced as $gq\rightarrow tZ$ (with $q = u, c$)
and top-quark-pair events, with one top quark
decaying through the $t \rightarrow Zq$ channel.
The analysis considers events with three leptons (electrons or muons), a $b$-tagged jet, possible
additional jets, and missing transverse momentum.
The data are found to be consistent with the background-only hypothesis
and 95\% confidence-level limits on the $t \rightarrow Zq$ branching ratios are set,
assuming only tensor operators of the Standard Model effective field theory
framework contribute to the $tZq$ vertices.
These are
$6.2 \times 10^{-5}$ ($13\times 10^{-5}$) for
$t\rightarrow Zu$ ($t\rightarrow  Zc$)
for a left-handed $tZq$ coupling, and
$6.6 \times 10^{-5}$ ($12\times 10^{-5}$)
in the case of a right-handed coupling.
These results are interpreted as 95\% CL upper limits
on the strength of corresponding couplings, yielding limits for
$|C_{uW}^{(13)*}|$ and $|C_{uB}^{(13)*}|$ ($|C_{uW}^{(31)}|$ and $|C_{uB}^{(31)}|$) of 0.15 (0.16),
and limits for $|C_{uW}^{(23)*}|$ and $|C_{uB}^{(23)*}|$ ($|C_{uW}^{(32)}|$ and $|C_{uB}^{(32)}|$) of 0.22 (0.21),
assuming a new-physics energy scale $\Lambda_\text{NP}$ of 1 TeV.
}
 
\author{The ATLAS Collaboration}
 
\AtlasRefCode{TOPQ-2019-06}
 
\AtlasDate{\today}

\hypersetup{pdftitle={ATLAS document},pdfauthor={The ATLAS Collaboration}}
 
\begin{document}
 
\maketitle
 
\tableofcontents

 
\section{Introduction}
\label{sec:intro}
The top quark is the heaviest elementary particle known and it decays almost exclusively
into \PW{}\Pqb~\cite{Zyla:2020zbs}.
In the Standard Model of particle physics (SM),
flavor-changing neutral-current (FCNC) processes involving a top
quark, an up-type quark and a \PZ{} boson are forbidden at tree level
and are strongly suppressed by the GIM mechanism~\cite{PhysRevD.2.1285}
at higher orders, leading to branching ratios
for top-quark decays via FCNC processes of the order of \SI{e-14}{}~\cite{Aguilar-Saavedra:2004mfd}.
However, several SM extensions predict such branching ratios to be between
\SI{e-4}{} and \SI{e-7}{}.
Examples of SM extensions are the quark-singlet model~\cite{Aguilar-Saavedra:2002phh},
the two-Higgs-doublet
model~\cite{Atwood:1996vj}, the Minimal Supersymmetric Standard Model
(MSSM)~\cite{Cao:2007dk}, the MSSM with R-parity
violation~\cite{Yang:1997dk}, models with warped extra
dimensions~\cite{Agashe:2006wa}, and extended mirror fermion
models~\cite{Hung:2017tts}.
 
FCNC couplings can be described by an effective field theory
(EFT)~\cite{Aguilar-Saavedra:2008nuh,Durieux:2014xla} that
extends the SM Lagrangian $\mathcal{L}_{\text{SM}}$ with
higher-dimensional operators suppressed by the scale of new physics,
$\Lambda_\text{NP}$, as shown in \cref{intro:eft}.
At order $\Lambda_\text{NP}^{-2}$ the strength of the anomalous couplings is
given by the Wilson coefficients $C_{k}$ that multiply dimension-six
operators $\mathcal{O}_{k}$,
 
\begin{equation}
\mathcal{L}_{\text{eff}} = \mathcal{L}_{\text{SM}} +\frac{1}{\Lambda_\text{NP}^2} \sum_k C_{k}\mathcal{O}_{k}.
\label{intro:eft}
\end{equation}
 
The relevant operators for an FCNC process with a top quark and a \PZ
boson, following the notation in Ref.~\cite{Grzadkowski:2010es},
are the operators $\mathcal{O}_{uB}^{(ij)}$ and $\mathcal{O}_{uW}^{(ij)}$
with $i \neq j$. 
The indices $i$ and $j$ of the operators refer to the flavor indices of
the quark generations. One index is always equal to 3 as a top quark
must be involved, while the other one is either 1 or 2, corresponding
to an up or charm quark.
The FCNC \tZq interactions can be introduced by vector and tensor
couplings, but only the latter are considered in this analysis because
they would produce most of the \enquote{FCNC-in-single-top-production}
signal~\cite{Durieux:2014xla}.
The FCNC operators can be left-handed (LH) or right-handed (RH).
The order of the indices $i$ and $j$ in \cref{intro:eft} defines the
chirality of the FCNC operators.
A linear combination of the $C_{uB}^{(13)}$ and $C_{uW}^{(13)}$ coefficients corresponds to the \tZu LH
coupling while a linear combination of the $C_{uB}^{(31)}$ and $C_{uW}^{(31)}$ coefficients defines the \tZu RH
coupling. Similarly, the \tZc couplings are defined by the $C_{uB}^{(23)}$ and $C_{uW}^{(23)}$
coefficients for the LH case, while the $C_{uB}^{(32)}$ and $C_{uW}^{(32)}$ coefficients
describe the RH case. For each linear combination, the two
coefficients assume the same value with an opposite sign~\cite{Durieux:2014xla}.
 
Experimental limits on the branching ratio of FCNC \ttoZq decays
were previously established by experiments
at the Large Electron--Positron Collider (LEP)~\cite{OPAL,ALEPH,L3,DELPHI},
the Hadron--Electron Ring Accelerator (HERA)~\cite{ZEUS},
the Tevatron~\cite{CDF,DZero}
and the Large Hadron Collider (LHC)~\cite{TOPQ-2014-08,TOPQ-2017-06,CMS-TOP-12-037-INT,CMS-TOP-12-039}.
The most stringent observed limits,
$\mathcal{B}(\ttoZu) < \SI{17e-5}{}$ and
$\mathcal{B}(\ttoZc) < \SI{24e-5}{}$~\cite{TOPQ-2017-06},
were set by ATLAS in a search for FCNC processes in \ttbar
decays only, using \SI{36.1}{\ifb} of $pp$ collision data at $\sqrt{s} = \SI{13}{\TeV}$.
The quoted limits apply to both the left- and right-handed couplings, as
the analysis is not sensitive to the chirality.
 
\begin{figure}[htb]
\centering
\begin{subfigure}[b]{.32\textwidth}
\includegraphics[width=1\textwidth]{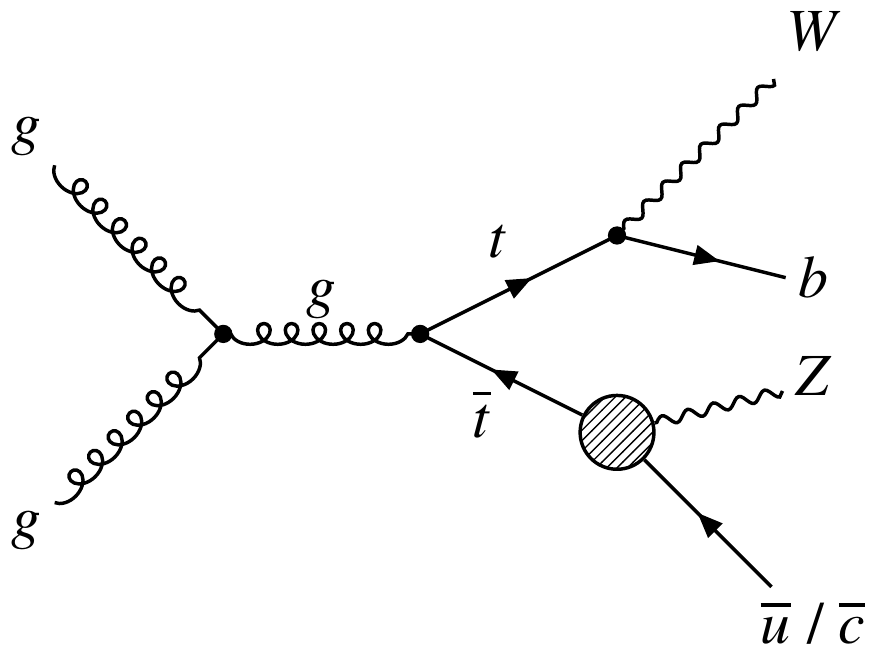}
\label{subfig:feynman-2}
\caption{ }
\end{subfigure}
\hspace{1.5cm}
\begin{subfigure}[b]{.32\textwidth}
\includegraphics[width=1\textwidth]{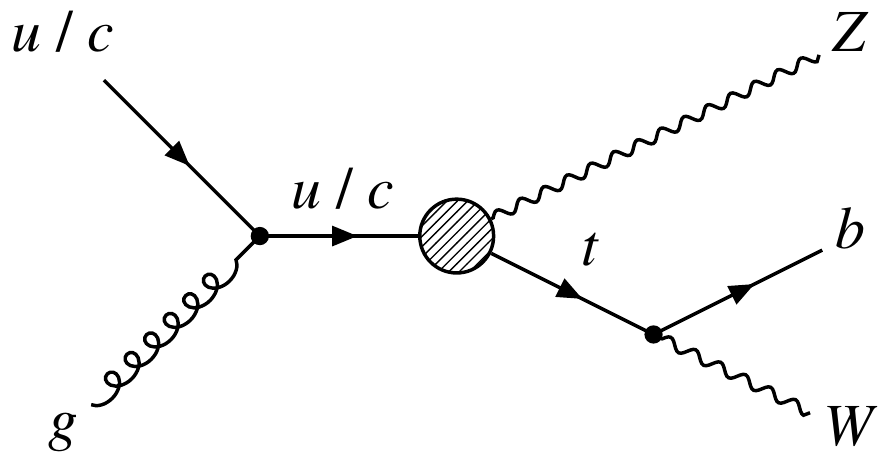}
\label{subfig:feynman-3}
\caption{ }
\end{subfigure}
\caption{Examples of the lowest-order Feynman diagrams for
(a) \ttbar production, with one top quark decaying through the
dominant mode in the SM and the other via an FCNC process and for
(b) single top-quark production via an FCNC process in the $s$-channel. }
\label{fig:feynman}
\end{figure}

This paper presents a search for FCNC \tZq couplings, using $pp$
collision data at $\sqrt{s} = \SI{13}{\TeV}$ collected by the ATLAS
experiment at the LHC and corresponding to an integrated luminosity of \lumi.
The search is performed by analyzing the top-quark decays in \ttbar
events as well as the production of single top quarks,
as illustrated in \Cref{fig:feynman}. In the former channel, one of the top quarks
decays through an FCNC process (\ttoZq) and the other through the
dominant mode (\ttoWb). In contrast, in the latter channel the production
of a single top quark proceeds through an FCNC process (\qtotZ).
Single top production with FCNC decay contributes negligibly
and is not considered in this analysis.
While single top-quark production gives the analysis more sensitivity to the FCNC
\tZu coupling, the \ttbar decay mode provides almost equal sensitivity to the
FCNC \tZu and \tZc couplings.
Since the FCNC production and decay processes are induced by the same
couplings, the production cross-section and decay branching ratio are
connected. Therefore, the FCNC single-top production cross-section can be
interpreted as the branching ratio of the corresponding FCNC decay. Thus,
the analysis results for the numbers of production and decay
signal events are translated into branching ratios for \ttoZq.
For both of the considered channels, only the trileptonic final state is selected,
in which the \PZ boson decays into charged leptons and the \PW boson
from the top quark decays leptonically.
The final states where either the \PZ boson or the \PW boson decays
hadronically are not considered because of the larger backgrounds.
Therefore, the analysis selects events with three leptons (electrons or muons), a \Pqb-tagged
jet, possible additional jets, and missing transverse momentum.
After the selection, the main background sources are diboson events and the productions of a single top quark in the $t$-channel or a \ttbar pair in association with a \PZ boson.
To improve the separation of signal from background events, a
multivariate technique is used, which was not employed in the previous analysis. The statistical analysis uses
a binned profile likelihood fit to the data.
 
\section{ATLAS detector}
\label{sec:atlas}
\newcommand{\AtlasCoordFootnote}{
ATLAS uses a right-handed coordinate system with its origin at the nominal interaction point (IP)
in the center of the detector and the $z$-axis along the beam pipe.
The $x$-axis points from the IP to the center of the LHC ring,
and the $y$-axis points upwards.
Cylindrical coordinates $(r,\phi)$ are used in the transverse plane,
$\phi$ being the azimuthal angle around the $z$-axis.
The pseudorapidity is defined in terms of the polar angle $\theta$ as $\eta = -\ln \tan(\theta/2)$.
Distances in the $\eta$--$\phi$ plane are measured in units of $\Delta R \equiv \sqrt{(\Delta\eta)^{2} + (\Delta\phi)^{2}}$.}

The ATLAS experiment~\cite{PERF-2007-01} at the LHC is a multipurpose particle detector
with a forward--backward symmetric cylindrical geometry and a near \(4\pi\) coverage in
solid angle.\footnote{\AtlasCoordFootnote}
It consists of an inner tracking detector surrounded by a thin superconducting solenoid
providing a \SI{2}{\tesla} axial magnetic field, electromagnetic and hadron calorimeters, and a muon spectrometer.
The inner tracking detector (ID) covers the pseudorapidity range \(|\eta| < 2.5\).
It consists of silicon pixel, silicon microstrip, and transition radiation tracking detectors.
Lead/liquid-argon (LAr) sampling calorimeters provide electromagnetic (EM) energy measurements
with high granularity.
A steel/scintillator-tile hadron calorimeter covers the central pseudorapidity range (\(|\eta| < 1.7\)).
The endcap and forward regions are instrumented with LAr calorimeters
for both the EM and hadronic energy measurements up to \(|\eta| = 4.9\).
The muon spectrometer surrounds the calorimeters and is based on
three large superconducting air-core toroidal magnets with eight coils each.
The field integral of the toroids ranges between \num{2.0} and \SI{6.0}{\tesla\metre}
across most of the detector.
The muon spectrometer includes a system of precision tracking chambers and fast detectors for triggering.
A two-level trigger system~\cite{TRIG-2016-01} is used to select events.
The first-level trigger is implemented in hardware and uses a subset of the detector information
to accept events at a rate below \SI{100}{\kHz}.
This is followed by a software-based trigger that
reduces the accepted event rate to \SI{1}{\kHz} on average
depending on the data-taking conditions.
An extensive software suite~\cite{ATL-SOFT-PUB-2021-001} is used in data simulation, in the reconstruction and
analysis of real and simulated data, in detector operations, and in the trigger and data acquisition systems of the experiment.


 
\section{Data and samples of simulated events}
\label{sec:datamc}
 
The data sample used in this analysis corresponds to \lumi of \pp
collisions at \(\sqrt{s} = \SI{13}{\TeV}\) collected by the ATLAS
detector during 2015--2018, after requiring stable LHC beams and that
all detector subsystems were operational~\cite{DAPR-2018-01}.
 
Candidate events were required to satisfy one of the single-electron
triggers or one of the single-muon triggers
~\cite{TRIG-2016-01,TRIG-2018-05,TRIG-2018-01}.
Single-lepton triggers with low transverse momentum (\pT) thresholds
and isolation requirements were combined in a logical OR with
higher-threshold triggers that had a looser identification criterion
and did not have any isolation requirement. The lowest \pT threshold
used for electrons was \SI[parse-numbers=false]{24}{\GeV}
(\SI[parse-numbers=false]{26}{\GeV}) in 2015 (2016--2018),
while for muons the corresponding threshold was \SI[parse-numbers=false]{20}{\GeV}
(\SI[parse-numbers=false]{26}{\GeV}).
 
To evaluate the effects of the detector resolution and acceptance on
the signal and background, and to estimate the SM backgrounds,
simulated event samples were produced using a \GEANT-based Monte
Carlo (MC) detector simulation~\cite{Agostinelli:2002hh-INT,SOFT-2010-01}.
Some of the samples used for evaluating systematic uncertainties did
not use the full \GEANT simulation but instead relied on
parameterized showers in the calorimeter~\cite{SOFT-2010-01}.
The top-quark mass in the event generators described below
was set to \mtop = \SI{172.5}{\GeV}.
In all samples, the decays of bottom and charm hadrons
were performed by \EVTGEN[1.2.0]~\cite{Lange:2001uf}, unless
stated otherwise.
 
The simulated data must account for the fact that significantly more
than one inelastic \pp collision occurs per bunch crossing.
The average number of collisions per bunch crossing ranged from 13 to 38
for the 2015--2018 data-taking periods.
Inelastic collisions were simulated using
\PYTHIA[8.186]~\cite{Sjostrand:2007gs} with the A3 set of tuned
parameters~\cite{ATL-PHYS-PUB-2016-017} and the
\NNPDF[2.3lo]~\cite{Ball:2012cx} set of parton distribution functions
(PDFs), and overlaid on the signal and background MC samples.
These simulated events were reweighted to match the conditions of the
collision data, specifically 
the number of additional \pp interactions in the same and neighboring bunch crossings (\pileup).
 
Several MC signal event samples were generated at next-to-leading order (NLO) in
QCD with \linebreak
\MGNLO[2.7.2]~\cite{Alwall:2014hca}, using the
\NNPDF[3.0nlo]~\cite{Ball:2014uwa} PDF set.
Parton showering and hadronization were modeled with
\PYTHIA[8.302] with the
\NNPDF[2.3lo] PDF set and
the A14 set of tuned parameters~\cite{ATL-PHYS-PUB-2014-021}.
Only events with leptonic decays (including $\tau$-leptons) of the \PW and \PZ bosons were
generated.
The TopFCNC Universal FeynRules Output (UFO)
\mbox{model~\cite{Alloul:2013bka,Degrande:2014tta,Durieux:2014xla}} was used
for the computation of top-quark FCNC production and decay processes
at NLO in QCD.
Since FCNC processes in both production and decay are considered in
this analysis, separate samples for each mode and for \tZu and \tZc
couplings were generated.
In order to study the chirality of these couplings,
separate samples with LH and RH couplings were
produced. 
 
Additional signal samples generated with the same version of \MGNLO were
interfaced to \HERWIG[7.1.6]~\cite{Bahr:2008pv,Bellm:2015jjp} instead of \PYTHIA[8.302]
to assess the uncertainty related to the choice of parton-shower model.
The \HERWIG[7.1] default set of tuned parameters~\cite{Bellm:2015jjp,Bellm:2017jjp}
was used together with the \MMHT[lo] PDF set~\cite{Harland-Lang:2014zoa}.
The decays of bottom and charm hadrons
were performed by \EVTGEN[1.7.0].
 
For the normalization, the branching ratios are set to the best observed
limits reported in \cref{sec:intro},
constraining $\mathcal{B}$($t~\to~q'W$) = 1 $-$ $\mathcal{B}$($t~\to~uZ/cZ$), with $q' = d, s, b$.
The FCNC \ttbar decay signal is normalized using the \ttbar
cross-section prediction at next-to-next-to-leading order (NNLO)
in QCD including the resummation of next-to-next-to-leading
logarithmic (NNLL) soft-gluon terms calculated using
\TOPpp[2.0]~\cite{Beneke:2011mq,Cacciari:2011hy,Baernreuther:2012ws,Czakon:2012zr,Czakon:2012pz,Czakon:2013goa,Czakon:2011xx}.
The FCNC single top-quark production signal normalization cross-section is calculated at
NLO using the TopFCNC model as implemented in \MGNLO.

The background is estimated using simulated samples
that contain at least two leptons and at least two jets. These samples
include the production of \ttbar, \ttH, \ttZ, \ttW, \SMtZq, \tW, \tWZ,
\Zjets, diboson, triboson, \threetops, \fourtops,
\ttWW, \ZH and \WH events.
 
The production of \ttbar and \ttH events was modeled using the
\POWHEGBOX[v2]~\cite{Frixione:2007nw,Nason:2004rx,Frixione:2007vw,Alioli:2010xd,Hartanto:2015uka}
generator at NLO with the \NNPDF[3.0nlo] PDF set
and the \hdamp parameter\footnote{The
\hdamp parameter is a resummation damping factor and one of the
parameters that controls the matching of \POWHEG matrix elements to
the parton shower and thus effectively regulates the
high-\pt radiation against which the \ttbar system recoils.} set
to 1.5\,\mtop for \ttbar~\cite{ATL-PHYS-PUB-2016-020} and
to $0.75 \times (2\,\mtop+ m_{\PH})$ for \ttH,
with $\mH = \SI{125}{\GeV}$. The events were interfaced
to \PYTHIA[8.230]~\cite{Sjostrand:2014zea} to model the parton shower,
hadronization, and underlying event, with parameters set according
to the A14 tune and using the \NNPDF[2.3lo]
set of PDFs. The decays of bottom and charm hadrons
were performed by \EVTGEN[1.6.0].
 
Additional \ttbar simulated samples are used to assess modeling
uncertainties~\cite{ATL-PHYS-PUB-2018-009}.
The impact of using a different parton shower and hadronization model is evaluated
by comparing the nominal \enquote{\POWPY} \ttbar sample with another event sample produced with the
\POWHEGBOX[v2] generator, but interfaced with \HERWIG[7.1.3],
which used the \HERWIG[7.1] default set of tuned parameters
and the \MMHT[lo] PDF set. To estimate the systematic uncertainty
in the choice of the \hdamp parameter, a sample generated in the same
way as the nominal one but with the \hdamp parameter set to 3.0\,\mtop
was produced.
 
The production of \ttZ\ and \ttW\ events was modeled using the
\MGNLO[2.3.3] generator at NLO with the \NNPDF[3.0nlo] PDF set.
The events were interfaced to \PYTHIA[8.210], which used the A14 tune and the
\NNPDF[2.3lo] PDF set.
 
Additional \ttZ simulated samples are used to assess modeling
uncertainties.
The impact of using a different parton shower and hadronization model is evaluated
by comparing the nominal \ttZ sample with an event sample produced with the
\MGNLO[2.6.2] generator interfaced with
\HERWIG[7.0.4],
which used the \HERWIG[7.0] default set of tuned parameters and the
\MMHT[lo] PDF set.
The decays of bottom and charm hadrons
were performed by \EVTGEN[1.6.0].
The uncertainty due to initial-state radiation (ISR) is estimated by
comparing the nominal event sample with two samples where the Var3c~\cite{ATL-PHYS-PUB-2014-021}
up and down variations of the A14 tune were employed.
 
The SM production of a single top quark in the $t$-channel in association with a \PZ boson
(\SMtZq) was modeled using the \MGNLO[2.3.3]
generator at NLO with the \NNPDF[3.0nlo] PDF set.
The events were interfaced with \PYTHIA[8.230], which used the A14 tune and the
\NNPDF[2.3lo] PDF set.
 
Similarly to \ttZ, additional \SMtZq simulated samples are used to assess modeling
uncertainties.
The impact of using a different parton shower and hadronization model is evaluated
by comparing the nominal \SMtZq sample with an event sample produced with the
\MGNLO[2.8.1] generator interfaced with \HERWIG[7.2.1], which used
the \HERWIG[7.1] default set of tuned parameters and the
\MMHT[lo] PDF set.
The decays of bottom and charm hadrons
were performed by \EVTGEN[1.7.0].
The uncertainty due to ISR is estimated by comparing the nominal \SMtZq sample with two additional samples,
which had the same settings as the nominal one, but employed the Var3c up and down variations of the A14 tune.
 
The associated production of a single top quark with a \PW boson (\tW)
was modeled by the \POWHEGBOX[v2]~\cite{Re:2010bp} generator at NLO
in QCD using the five-flavor scheme and the \NNPDF[3.0nlo] set of PDFs.
The diagram removal (DR) scheme~\cite{Frixione:2008yi} was used to
remove interference and overlap with \ttbar production.
The events were interfaced to \PYTHIA[8.230], which used the A14
tune and the \NNPDF[2.3lo] set of PDFs.
 
The production of \tWZ\ events was modeled using the \MGNLO[2.3.3]
generator at NLO with the \NNPDF[3.0nlo] PDF set.
The events were interfaced with \PYTHIA[8.212], which used the A14 tune and the
\NNPDF[2.3lo] PDF set.
The DR scheme was employed to handle the interference between the \tWZ and
\ttZ processes.
A sample with an alternative scheme described in
Ref.~\cite{Demartin:2016axk} was produced to assess the associated systematic uncertainty.
 
The \POWHEGBOX[v1] MC generator~\cite{Alioli:2008gx}
was used to simulate at NLO accuracy
the hard-scattering
processes of \PZ boson production and decay in the electron, muon, and $\tau$-lepton
channels. It was interfaced to \PYTHIA[8.186]
for the modeling of the parton shower, hadronization, and underlying
event, with parameters set according to the \AZNLO
tune~\cite{STDM-2012-23}. The \CT[10nlo]~\cite{Lai:2010vv} PDF set was used
for the hard-scattering processes, whereas the \CTEQ[6L1]~\cite{Pumplin:2002vw} PDF
set was used for the parton shower. The effect of
QED final-state radiation was simulated with \PHOTOSpp[3.52]~\cite{Golonka:2005pn,Davidson:2010ew}.
 
Samples of diboson final states (\VV, with $V$ = \PW, \PZ) were simulated with the
\SHERPA[2.2.1] or 2.2.2~\cite{Bothmann:2019yzt} generator depending on the process,
including off-shell effects and Higgs boson contributions where appropriate.
Fully leptonic final states and semileptonic final states, where one boson
decays leptonically and the other hadronically, were generated using
matrix elements at NLO accuracy in QCD for up to one additional parton
and at LO accuracy for up to three additional parton
emissions. Samples for the loop-induced processes $gg \to VV$ were
generated using LO-accurate matrix elements for up to one
additional parton emission for both the cases of fully leptonic and
semileptonic final states. The matrix element calculations were matched
and merged with the \SHERPA parton shower based on Catani--Seymour
dipole factorization~\cite{Gleisberg:2008fv,Schumann:2007mg} using the MEPS@NLO
prescription~\cite{Hoeche:2011fd,Hoeche:2012yf,Catani:2001cc,Hoeche:2009rj}.
The virtual QCD corrections were provided by the
\OPENLOOPS library~\cite{Buccioni:2019sur,Cascioli:2011va,Denner:2016kdg}. The
\NNPDF[3.0nnlo] set of PDFs was used, along with the
dedicated set of tuned parton-shower parameters developed by the
\SHERPA authors.
Electroweak production of a diboson in association with two jets
($VVjj$) was simulated with the \SHERPA[2.2.2]
generator. The LO-accurate matrix elements were matched to a parton
shower based on Catani--Seymour dipole factorization using the MEPS@LO
prescription. Samples were generated using the \NNPDF[3.0nnlo] PDF set,
along with the dedicated set of tuned parton-shower parameters
developed by the \SHERPA authors.
The decays of bottom and charm hadrons are performed with built-in \SHERPA features.
An invariant mass of $m_{\ell\ell} >$ \SI{4}{\GeV} was required
at matrix-element level for any pair of same-flavor charged leptons.
 
To assess the uncertainty that the generator contributes to the simulation of diboson final states, alternative samples are
employed. For these, the \POWHEGBOX[v2]~\cite{Nason:2013ydw} generator
was used instead of \SHERPA. The effect of singly resonant
amplitudes and interference effects due to $Z/\gamma^*$ and
same-flavor lepton combinations in the final state were included where appropriate.
Interference effects between $WW$ and $ZZ$ for same-flavor charged leptons and
neutrinos were ignored.
Events were interfaced to \PYTHIA[8.186]
for the modeling of the parton shower, hadronization, and underlying
event, with parameters set according to the \AZNLO
tune. The \CT[10] PDF set was used
for the hard-scattering processes, whereas the \CTEQ[6L1] PDF
set was used for the parton shower.
The factorization and renormalization scales were set to the invariant mass of
the boson pair. The same invariant mass selection as for the \SHERPA
samples was applied.
 
The production of triboson (\VVV, with $V$ = \PW, \PZ) events was simulated with the
\SHERPA[2.2.2] generator.
Matrix elements, accurate to NLO for the inclusive process and to LO for up to
two additional parton emissions, were matched and merged with the \SHERPA parton
shower based on Catani--Seymour dipole factorization
using the \MEPSatNLO prescription.
The virtual QCD corrections for matrix elements at NLO accuracy were
provided by the \OPENLOOPS library.
Samples were generated using the \NNPDF[3.0nnlo] PDF set, along with
the dedicated set of tuned parton-shower parameters developed by the \SHERPA authors.
The decays of bottom and charm hadrons are performed with built-in \SHERPA features.
 
The production of \fourtops events was modeled using the \MGNLO[2.6.2]
generator at NLO with the \NNPDF[3.1nlo]~\cite{Ball:2014uwa}  PDF set.
The events were interfaced with \PYTHIA[8.230], which used the A14 tune and the
\NNPDF[2.3lo] PDF set.
The decays of bottom and charm hadrons were simulated using the
\EVTGEN[1.6.0] program.
 
Other rare top-quark processes, namely the production of \ttWW and \threetops events, were modeled using the
\MGNLO generator at LO interfaced with \PYTHIA[8], which used the A14 tune.
The associated production of a Higgs boson with a \PW or \PZ
boson, \VH, was modeled using \PYTHIA[8.186]~ with the A14 tune and
the \NNPDF[2.3lo] PDF set.
 
\LDnote{}{Need to add VH}
 
Throughout the paper the various MC samples are merged or split as
follows.
The \ttZ and \tWZ backgrounds are combined.
The diboson contribution is split according to the origin of the
associated jets using generator-level information. Their origin is
determined by matching, within a cone of size $\Delta R = 0.3$, jets to hadrons
with $\pt > \SI{5}{\GeV}$. If one of the jets
contains a \Pqb- or \Pqc-hadron, then it is classified as diboson +
heavy flavor (\VVHF), otherwise the event is classified as diboson +
light flavor (\VVLF).
The \ttbar, \tW, \Zjets, \VV and \ttV processes with
two prompt\footnote{Prompt leptons are leptons from the decay of \PW or \PZ bosons, either
directly or through an intermediate $\tau\to \ell\nu\nu$ decay, or from the
semileptonic decay of top quarks.} leptons and one nonprompt or fake lepton (a jet misidentified as a lepton) are shown together
and called \enquote{Fakes}.
The other minor backgrounds, namely \ttW, \ttH, \VH, \ttWW, triboson, \threetops
and \fourtops, are merged and called \enquote{Other bkg.}.


 
\section{Object reconstruction}
\label{sec:reco}
 
The reconstruction of the basic objects used in the analysis is
described in the following.
The primary vertex~\cite{ATL-PHYS-PUB-2015-026} is selected as the \pp
vertex candidate with the highest sum of the squared transverse
momenta of all associated tracks with $\pT > \SI{500}{\MeV}$.

Electron candidates are reconstructed from energy clusters in the
EM calorimeter that match a reconstructed track~\cite{EGAM-2018-01}.
The clusters are required to be within the range $\abseta < 2.47$,
excluding the transition region between the barrel and endcap
calorimeters at $1.37 < \abseta < 1.52$.
Each electron candidate's transverse impact parameter relative to the beam axis, $d_0$,
divided by its estimated uncertainty must satisfy $|d_0|/\sigma(d_0) < 5$, while the
longitudinal distance $z_0$ from the reconstructed primary vertex to the point where
$d_0$ is measured must satisfy $|z_0\sin(\theta)| < \SI{0.5}{\milli\metre}$.
Electron candidates must also satisfy a transverse momentum requirement
of $\pt > \SI{15}{\GeV}$.
A likelihood-based discriminant is constructed from a set of
variables that enhance the electron selection, while
rejecting photon conversions and hadrons misidentified as
electrons.
An $\eta$- and \pt-dependent selection on the likelihood discriminant
is applied, and the \enquote{Medium} identification~\cite{EGAM-2018-01} is used.
Electrons are also required to be isolated using criteria based on
ID tracks.
Nonprompt leptons are rejected using a boosted decision tree
(BDT) discriminant based on isolation and \Pqb-tagging variables,
referred to as the nonprompt-lepton BDT~\cite{HIGG-2017-02}.
The efficiency at the chosen working point for electrons satisfying
the isolation criteria is about
\SI{70}{\%} for a \pT of \SI{20}{\GeV} and reaches a plateau of
\SI{95}{\%} at a \pT of \SI{100}{\GeV}. The corresponding rejection factor for leptons
from the decay of \Pqb-hadrons is about 50, estimated from a simulated \ttbar sample.
Correction factors are applied to simulated electrons to take into
account the small differences in trigger, reconstruction, identification and
isolation efficiencies between data and MC simulation.
\LDnote{}{Need to check values for PLV}

Muon candidates are reconstructed by combining a reconstructed track
from the inner detector with one from the muon
spectrometer, and are required to
have $\pt>\SI{15}{\GeV}$ and $\abseta<2.5$ and to meet the
\enquote{Medium} identification~\cite{MUON-2018-03-INT} criteria.
Similarly to electrons, muon candidates must have
$|d_0|/\sigma(d_0) < 3$ and $|z_0\sin(\theta)| < \SI{0.5}{\milli\metre}$.
To reject misidentified muon candidates, several quality requirements are imposed on the
muon candidate.
An isolation requirement based on ID tracks is imposed, and a
threshold is set for the nonprompt-lepton BDT output.
The efficiency at the chosen working point for muons satisfying
the isolation criteria is about
\SI{80}{\%} for a \pT of \SI{20}{\GeV} and reaches a plateau of
\SI{99}{\%} at a \pT of \SI{100}{\GeV}. The corresponding rejection factor for leptons
from the decay of \Pqb-hadrons is about 20, estimated from a simulated \ttbar sample.
Like for electrons, correction factors are applied to simulated muons
to account for the small differences between data and simulation.
\LDnote{}{Need to check values for PLV}

Jets are reconstructed from the particle-flow
objects~\cite{PERF-2015-09} using the \antikt
algorithm~\cite{Cacciari:2008gp,Fastjet} with the radius parameter set to $R=0.4$.
Their calibration follows the methodology described in Ref.~\cite{JETM-2018-05}.
Jets are required to have $\pT > \SI{25}{\GeV}$ and $\abseta < 2.5$.
To suppress jets arising from \pileup, a discriminant called the
\enquote{jet vertex tagger} (JVT) is constructed using a
two-dimensional likelihood method~\cite{PERF-2014-03}.
The jet energy scale and resolution are corrected with $\eta$- and
\pT-dependent scale factors.

To identify jets containing a \Pqb-hadron (\Pqb-jets), the
\enquote{DL1r} multivariate algorithm
is employed~\cite{FTAG-2019-07-INT}.
It uses impact parameter and secondary and tertiary vertex
information from tracks contained in the jet as input.
Operating points are defined by a threshold value for the \btag discriminant
output and are chosen to provide a specific \Pqb-jet
efficiency in an inclusive \ttbar sample.
Candidate \Pqb-jets must have a \btag discriminant value that exceeds
a threshold corresponding to a \SI{70}{\%} \Pqb-jet selection efficiency.
With this criterion, \SI{0.25}{\%} of light-jets, containing neither a \Pqb- nor a \Pqc-hadron, are misidentified as \Pqb-jets, as are \SI{10}{\%} of jets initiated by \Pqc-quarks.
Correction factors are derived and applied to correct for differences
in \Pqb-jet selection efficiency and the mistagging rates between data
and MC \mbox{simulation~\cite{FTAG-2019-07-INT}}.

The missing transverse momentum, with magnitude \MET, is calculated as
the negative of the vector sum of the transverse momenta of all
reconstructed objects.
To account for soft hadronic activity, a term including tracks
associated with the primary vertex but not with any of the
reconstructed objects is added to the \MET calculation~\cite{PERF-2016-07,ATLAS-CONF-2018-023}.

To avoid cases where the detector response to a single physical
object is reconstructed as two separate final-state objects, an
overlap removal procedure is used.
If electron and muon candidates share a track, the electron candidate
is removed. After that, if the $\DeltaR_{y,\phi}$
distance\footnote{$\DeltaR_{y,\phi}$ is the Lorentz-invariant distance in the
rapidity--azimuthal-angle plane, defined as $\DeltaR_{y,\phi}
=\sqrt{(\Delta y)^2+(\Delta\phi)^2}$, where $y$ is the rapidity,
defined as $y = (1/2)$ ln $[(E+p_z)/(E-p_z)]$.}
between a jet and an electron candidate is less than 0.2, the jet is
discarded. If multiple jets satisfy this requirement, only the closest
jet is removed.
For jet--electron distances between 0.2 and 0.4, the electron
candidate is removed.
If the distance between a jet and a muon candidate is less than 0.2,
and the jet has less than three associated tracks, the jet is
removed. Any muon subsequently found at a distance of less than 0.4 from a
jet is removed.


 
\section{Event reconstruction and selection}
\label{sec:evtsel}
 
The analysis searches for effects of FCNC \tZq couplings both in \ttbar decay and
in single-top-quark production processes.
In the first process, one of the top quarks decays through the
dominant mode into a \PW boson and a \Pqb-quark
(hereafter called the \enquote{SM top quark}, denoted by $t_{\mathrm{SM}}$),
while the other top quark (hereafter called the \enquote{FCNC top quark}, denoted by $t_{\mathrm{FCNC}}$)
is assumed to decay into a \PZ boson and a $u$- or $c$-quark.
In the second process, the production of a single top quark is assumed to proceed
through an FCNC interaction in association with a \PZ boson, while its
decay is through the dominant mode.
In each channel, only the trilepton final state is targeted, in
which the \PZ and \PW bosons decay leptonically.
Therefore, the final state of the FCNC process in \ttbar decays is
characterized by the presence of three leptons,
at least two jets, one of which is a \Pqb-jet, and
missing transverse momentum from the escaping neutrino.
The final state of the FCNC process in single top-quark production
is instead characterized by the presence of three leptons, a \Pqb-jet,
up to one additional jet,
and missing transverse  momentum.
Due to the different final states, two separate signal regions (SRs)
are defined, targeting the two processes: SR1 targets FCNC processes
in \ttbar decays while SR2 targets FCNC processes in single top-quark
production. The SRs share common selections
for the leptons and they differ in their top-quark reconstruction and
jet multiplicity requirements.
 
In both SRs, exactly three leptons (electrons or muons) that do not
all have the same charge are required.
One of the leptons must
have $\pt > \SI{27}{\GeV}$, because of the trigger thresholds, and
must be matched, with $\Delta R < 0.15$, to the lepton reconstructed
by the trigger.
Events with a fourth reconstructed lepton with $\pt > \SI{15}{\GeV}$ are vetoed.
At least one opposite-sign same-flavor lepton pair
(OSSF) with an invariant mass in the range
$|\mll - \SI{91.2}{\GeV}| < \SI{15}{\GeV}$ is required.
In the \mee and \emm channels the pair is uniquely identified, whereas
in the \eee and \mmm channels both of the possible combinations are considered and the
pair with the invariant mass closer to the \PZ boson mass is chosen.
The lepton not used to reconstruct the \PZ boson is assumed to be the
one coming from the \PW boson, $\ell_{\PW}$.
In SR2, to help reject background sources
with a third nonprompt lepton, events are required to have
$\mT(\ell_{\PW},\Pgn) > \SI{40}{\GeV}$.\footnote{The transverse mass is calculated
using the momentum of the lepton associated with the \PW boson, the \MET and
the azimuthal angle, $\phi$, between them:
$\mT$($\ell_{\PW},\Pgn$)$ = \sqrt{2\pT^{\ell}\MET \left(1-\cos\Delta\phi\right)}$.}
 
In SR1 the selected events have at least two jets, with exactly one \Pqb-tagged.
In SR2 the selected events have one or two jets, with exactly one \Pqb-tagged.
For events with exactly two jets, orthogonality between SR1 and SR2 is
ensured by using an invariant mass cut on reconstructed top-quark
candidates, as defined in the following.
An additional SR targeting the FCNC \tZc coupling in \ttbar decay, based on
the presence of a \Pqc-jet, was considered. The \Pqc-tagging was done
using the soft-muon tagging technique employed in Ref.~\cite{TOPQ-2017-17-INT}.
With the current dataset, this SR was found to bring only marginal
improvements to the final limits.
 
In the events having at least two jets with one of them being
\Pqb-tagged, the reconstruction of FCNC and SM top-quark candidates is
based on the \enquote{FCNC-in-\ttbar-decay} signal hypothesis.
The kinematics of the top-quark candidates are reconstructed from
the corresponding decay particles by minimizing the following expression:
 
\begin{equation}
\chi^2_{\ttbar}  =  \frac{\left(m^{\mathrm{reco}}_{j_a\ell\ell}-m_{t_{\mathrm{FCNC}}}\right)^2}{\sigmaFCNC^2}+
\frac{\left(m^{\mathrm{reco}}_{j_b\ell_{\PW}\nu}-m_{t_{\mathrm{SM}}}\right)^2}{\sigmaSM^2}
+ \frac{\left(m^{\mathrm{reco}}_{\ell_{\PW}\nu}-m_{\PW}\right)^2}{\sigma_{\PW}^2},
\label{eq:chi2tt}
\end{equation}
 
where $m^{\mathrm{reco}}_{j_a\ell\ell}$,
$m^{\mathrm{reco}}_{j_b\ell_{\PW}\nu}$, and $m^{\mathrm{reco}}_{\ell_{\PW}\nu}$ are
the reconstructed masses of the \PZ{}\Pq, \PW{}\Pqb, and $\ell_{\PW{}}\nu$ systems, respectively.
The minimization has two independent parts. The first is the jet permutation,
where any non-\Pqb-tagged jet can be assigned to $j_a$,
while $j_b$ must correspond to a \Pqb-tagged jet. The second is
the minimization of the $\chi^2_{\ttbar}$ for each permutation by
varying the longitudinal component of the neutrino momentum, $p^{\nu}_z$,
to determine the most probable value while its transverse component is set to the missing transverse
momentum in the event.
 
This procedure assigns a reconstructed jet to the \Pq-quark from the decay of the
FCNC top quark and  determines the $p^{\nu}_z$ value to reconstruct the four-momenta of the two top-quark candidates.
 
In~\cref{eq:chi2tt}, the central values ($m_{t_{\mathrm{FCNC}}}$,
$m_{t_{\mathrm{SM}}}$ and $m_{\PW}$) and the widths ($\sigmaFCNC$,
$\sigmaSM$ and $\sigma_{\PW}$) of the distributions of the
reconstructed masses of the top quark and \PW boson candidates are
taken from reconstructed
simulated FCNC-in-\ttbar-decay signal events
that undergo the common object selection procedure just described.
This is done by matching the true \Pq- and  \Pqb-quarks in the simulated events to the reconstructed jets,
setting the longitudinal momentum of the neutrino to the $p_z$ of the true generated neutrino,
and the transverse component to the missing transverse momentum in the
event, and then performing a likelihood fit with a Bukin
function\footnote{These fits use a generalization of the Gaussian
function to allow for asymmetric tails in the distribution.
The overall normalization is fixed to the yield and the shape of the
function is determined by five parameters: the peak
position, the width of the core, the asymmetry, the size of the lower
tail, and the size of the higher tail. From these parameters, only the
peak position and the width enter the $\chi^2$.}~\cite{Bukin}
to the masses of the reconstructed top quarks and \PW boson.
The mass values for the LH coupling are reported in \cref{tab:bukin}.
Compatible mass values are obtained for the RH coupling.
The fraction of reconstructed top-quark candidates that are matched to
the true simulated particles within a cone of size $\Delta R = 0.4$ is
$\epsilon_{t_{\mathrm{FCNC}}} = \SI{75}{\%}$ for the FCNC top-quark candidates
and $\epsilon_{t_{\mathrm{SM}}} = \SI{54}{\%}$ for the SM top-quark candidates,
where the difference comes from the fact that for the SM top-quark
decay the match of the missing transverse momentum with the generated neutrino is less efficient.
 
\begin{table}[htbp]
\caption{
Summary of the mean values and standard deviations
of the invariant mass distributions for the top-quark
candidates and the \PW boson.
These values are obtained from the Bukin fits using the FCNC-in-\ttbar-decay signal samples with the LH coupling.
The two FCNC \tZu and \tZc coupling samples are combined.
}
\label{tab:bukin}
\centering
\small
\begin{tabular}{lcccccc}
\toprule
& \multicolumn{2}{c}{FCNC top quark} & \multicolumn{2}{c}{SM top quark} & \multicolumn{2}{c}{\PW boson} \\
& $m_{t_{\mathrm{FCNC}}}$ [GeV] & $\sigmaFCNC$ [GeV] & $m_{t_{\mathrm{SM}}}$ [GeV] & $\sigmaSM$ [GeV] & $m_{\PW}$ [GeV] & $\sigma_{\PW}$ [GeV] \\
\midrule
FCNC in \ttbar decay (LH) & 171.0 & 11.1 & 166.5 & 23.2 & 80.5 & 15.4 \\
\bottomrule
\end{tabular}
\end{table}
 
Under the FCNC-in-single-top-quark-production signal hypothesis, the
SM top-quark candidate is instead reconstructed in events having one or two
jets, with exactly one $b$-tagged.
The missing transverse momentum is assumed to be the transverse component of the neutrino momentum,
while the most probable value of $p^{\nu}_z$ is determined by minimizing the following expression:
 
\begin{equation}
\chi^2_{\Pqt\PZ}  =
\frac{\left(m^{\mathrm{reco}}_{j_{b}\ell_{\PW}\nu}-m_{t_{\mathrm{SM}}}\right)^2}{\sigmaSM^2}
+ \frac{\left(m^{\mathrm{reco}}_{\ell_{\PW}\nu}-m_{\PW}\right)^2}{\sigma_{\PW}^2},
\label{eq:chi2tZ}
\end{equation}
 
where $m^{\mathrm{reco}}_{j_b\ell_{\PW}\nu}$ and $m^{\mathrm{reco}}_{\ell_{\PW}\nu}$ are
the reconstructed masses of the \PW{}\Pqb and $\ell_{\PW}\nu$ systems, respectively.
In~\cref{eq:chi2tZ}, the central values for the masses and widths
of the top quark and \PW boson are taken from reconstructed
simulated FCNC-in-\ttbar-decay signal events, as is done in~\cref{eq:chi2tt}.\footnote{Using the central values for the masses and widths extracted from the
FCNC single-top production signal sample does not have a significant effect on the final results.}
Therefore, in the events with two jets,
the four-momentum of the SM top-quark candidate reconstructed under the
FCNC-in-single-top-quark-production signal hypothesis is the same as that
reconstructed under the FCNC-in-\ttbar-decay signal hypothesis.
In this case, the fraction of reconstructed top-quark candidates that
are matched to the true simulated particles within a cone of size
$\Delta R = 0.4$ is $\epsilon_{t_{\mathrm{SM}}} = \SI{71}{\%}$.
 
In SR1, the mass of the FCNC top-quark candidate, \mtopfcnc, is required to be
within 2\sigmaFCNC of \mtopvalue, while no requirement is placed on the
mass of the SM top-quark candidate, \mtopsm.
In SR2, the mass of the SM top-quark candidate is required to be within 2\sigmaSM of \mtopvalue.
In addition, to ensure orthogonality with SR1, for events with exactly two jets
the mass of the FCNC top-quark candidate
is required to be more than 2\sigmaFCNC from \mtopvalue.
\Cref{tab:cutsummary} summarizes the selection criteria applied to
the signal regions considered.
With these criteria, 496 data events are selected in SR1 and 460
are selected in SR2.
 
\begin{table}[htbp]
\caption{
Overview of the requirements applied to select the events in the signal regions.
OSSF is an opposite-sign same-flavor lepton pair,
$m_{\PZ}$ = \SI{91.2}{\GeV} and $m_{\Pqt}$ = \mtopvalue.
}
\label{tab:cutsummary}
\centering
\begin{tabular}{ccc}
\toprule
\multicolumn{3}{c}{Common selections} \\
\midrule
\multicolumn{3}{c}{Exactly 3 leptons with $\pT(\ell_1)> \SI{27}{\GeV}$} \\
\multicolumn{3}{c}{$\ge$ 1 \OSSF pair, with $|\mll - m_{\PZ}| < \SI{15}{\GeV}$} \\
\toprule
SR1 & \multicolumn{2}{c}{SR2} \\
\midrule
$\ge$ 2 jets & 1 jet & 2 jets \\
1 \Pqb-jet  & 1 \Pqb-jet & 1 \Pqb-jet \\
-- & $\mT$($\ell_{\PW},\Pgn$)$ > \SI{40}{\GeV}$ & $\mT$($\ell_{\PW},\Pgn$)$ > \SI{40}{\GeV}$ \\
$|m^{\mathrm{reco}}_{j_a\ell\ell} - m_{\Pqt}| < 2\sigmaFCNC$ & -- & $|m^{\mathrm{reco}}_{j_a\ell\ell} - m_{\Pqt}| > 2\sigmaFCNC$ \\
-- & $|m^{\mathrm{reco}}_{j_b\ell_W\nu} - m_{\Pqt}| < 2\sigmaSM$ & $|m^{\mathrm{reco}}_{j_b\ell_W\nu} - m_{\Pqt}| < 2\sigmaSM$ \\
\bottomrule
\end{tabular}
\end{table}
 
\begin{figure}[htbp]
\centering
\begin{subfigure}[b]{.45\textwidth}
\includegraphics[width=1\textwidth]{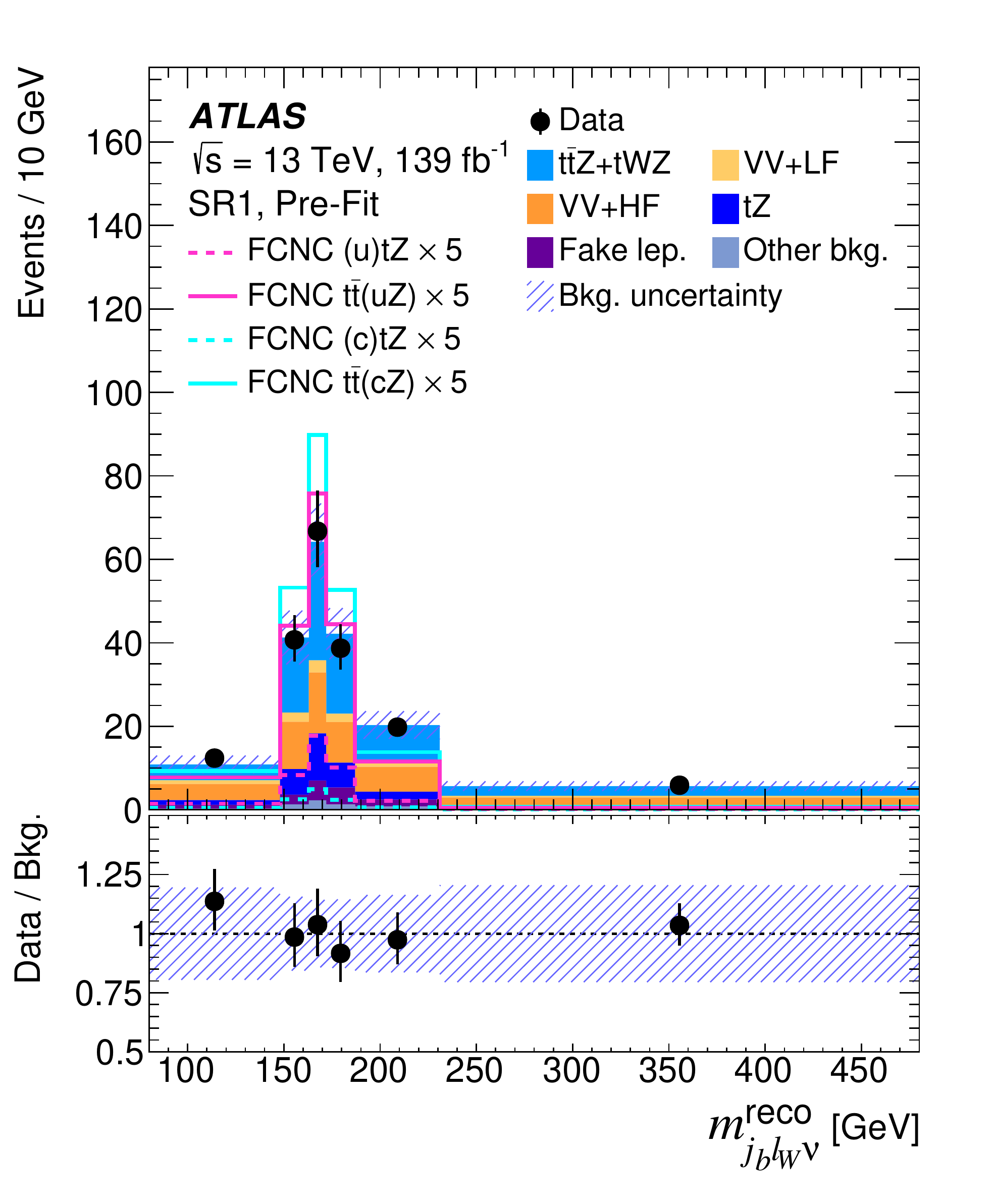}
\label{subfig:eventsel:tzu:unblinded:prefit-1}
\caption{ }
\end{subfigure}
\begin{subfigure}[b]{.45\textwidth}
\includegraphics[width=1\textwidth]{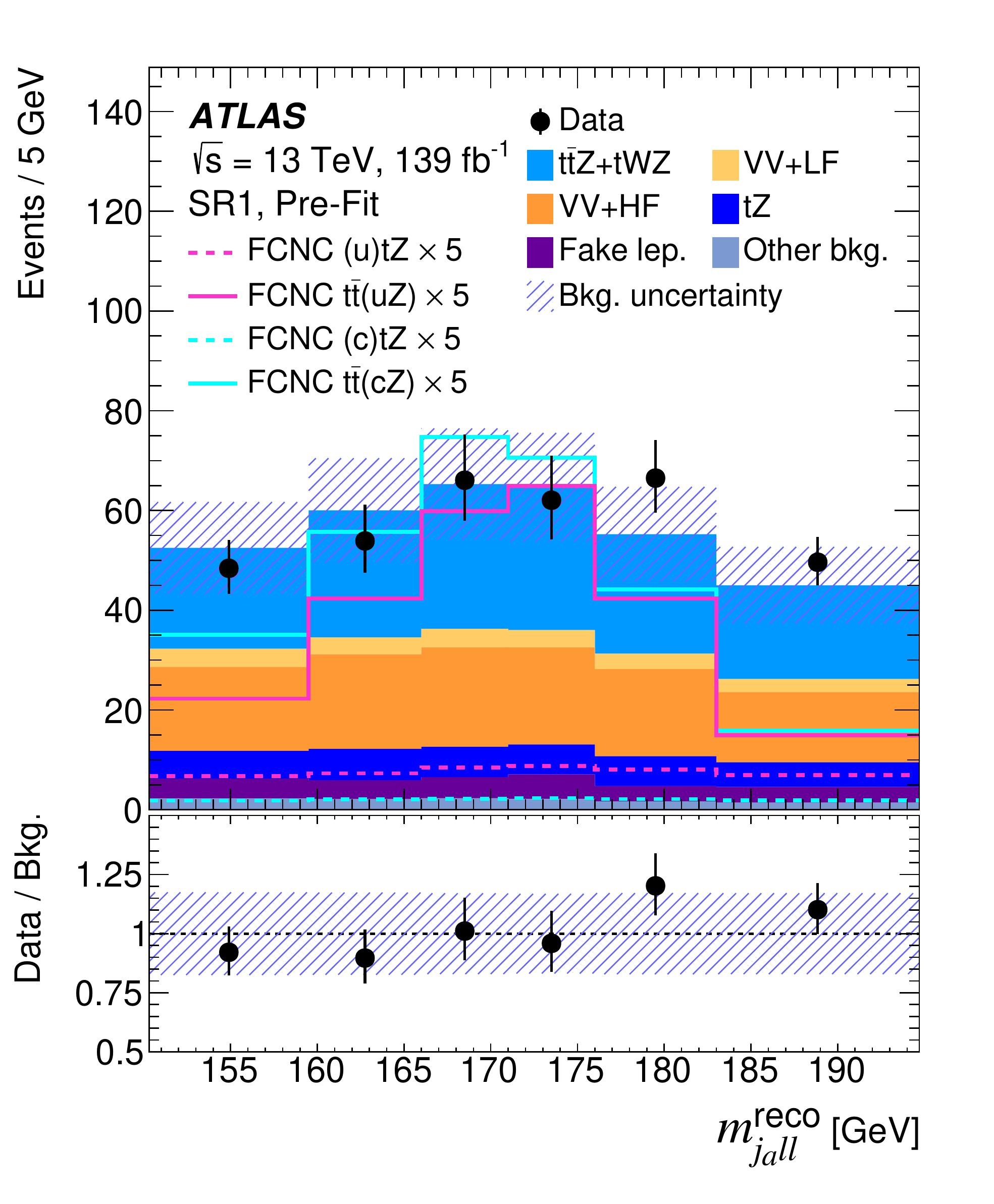}
\label{subfig:eventsel:tzu:unblinded:prefit-2}
\caption{ }
\end{subfigure}
\begin{subfigure}[b]{.45\textwidth}
\includegraphics[width=1\textwidth]{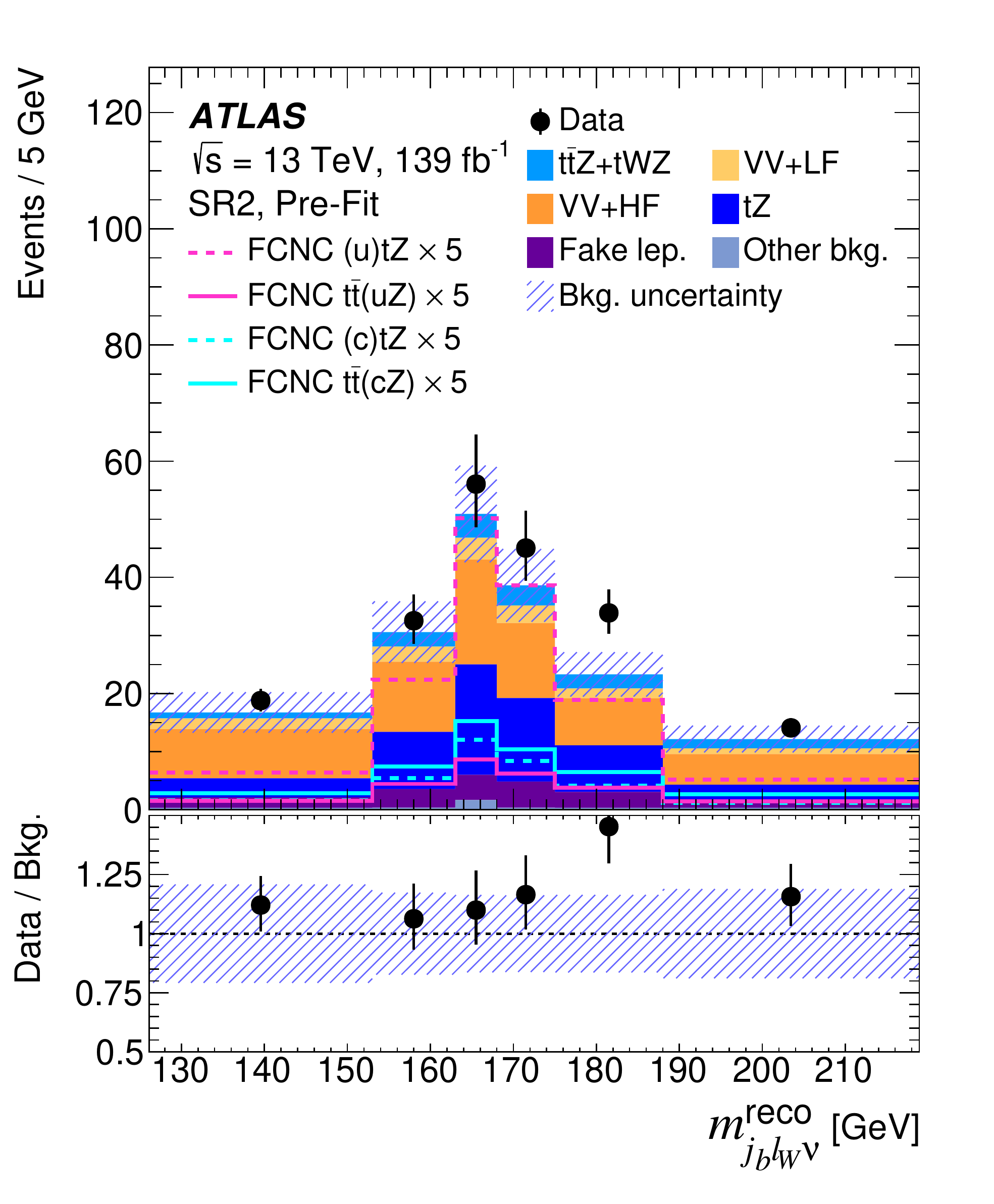}
\label{subfig:eventsel:tzu:unblinded:prefit-3}
\caption{ }
\end{subfigure}
\begin{subfigure}[b]{.45\textwidth}
\includegraphics[width=1\textwidth]{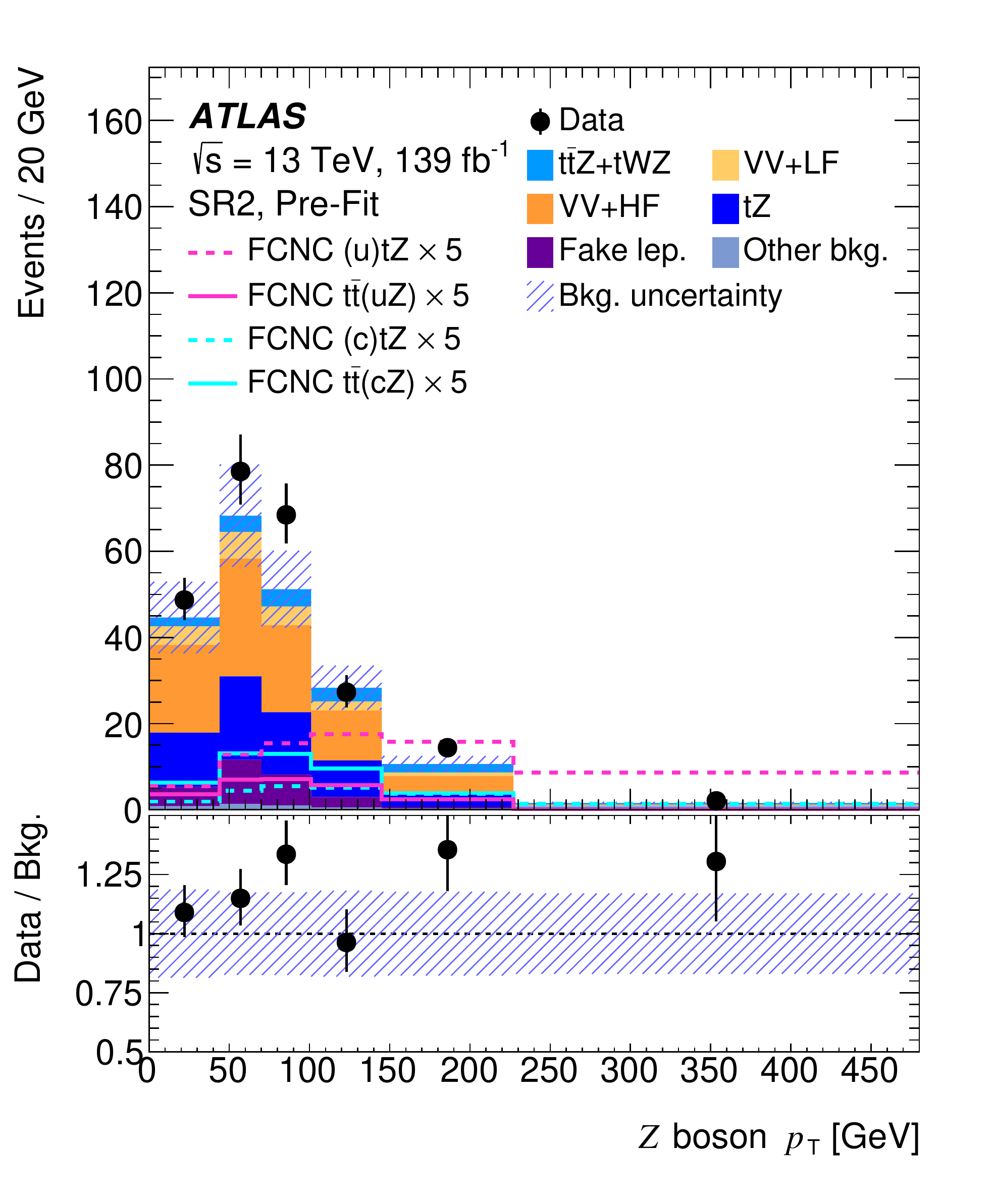}
\label{subfig:eventsel:tzu:unblinded:prefit-4}
\caption{ }
\end{subfigure}
\caption{Comparison between data and background prediction
before the fit (\enquote{Pre-Fit})
for some kinematic distributions in the SRs.
The distributions are:
(a) the mass of the SM top-quark candidate in SR1,
(b) the mass of the FCNC top-quark candidate in SR1,
(c) the mass of the SM top-quark candidate in SR2 and
(d) the transverse momentum of the \PZ boson candidate in SR2.
The uncertainty band includes both the statistical and systematic
uncertainties in the background prediction.
The four FCNC LH signals are also shown separately, normalized to
five times the cross-section corresponding to the most stringent observed
branching ratio limits~\cite{TOPQ-2017-06}.
The first (last) bin in all distributions includes the underflow
(overflow).
The lower panels show the ratios of the data (\enquote{Data})
to the background prediction (\enquote{Bkg.}).
}
\label{fig:eventsel:tzu:unblinded:prefit}
\end{figure}

\Cref{fig:eventsel:tzu:unblinded:prefit} shows the distributions of
the masses of the two top-quark candidates in SR1, and the mass of the
top-quark candidate and the \pT of the reconstructed \PZ boson
in SR2. These kinematic distributions are some of the key features that
distinguish signal events from the backgrounds and they are utilized in
the multivariate analysis described in \cref{sec:mva}.
In SR1, the dominant signal is the FCNC-in-\ttbar-decay events
(shown with solid lines in \Cref{fig:eventsel:tzu:unblinded:prefit}
separately for the \tZu and \tZc couplings),
while the FCNC-in-single-top-quark-production contribution (shown with
dashed lines) is smaller. In contrast, SR2 is more sensitive to the \tZu
FCNC-in-single-top-quark-production signal, with similar smaller contributions
from the other three signals.
After the event selection the main background sources are \ttZ, \SMtZq and diboson production.
 
\FloatBarrier

 
\section{Background estimation and separation from signal}
\label{sec:mva}

Two classes of backgrounds are considered: processes in which three or
more prompt leptons are produced, such as diboson production or the
associated production of top quarks (\ttZ, \tWZ, \SMtZq, \ttW, \ttH)
and processes with two prompt leptons in the final state along with
one additional nonprompt or fake lepton that satisfies the selection criteria, such as \ttbar, \tW, and \Zjets.
Such nonprompt or fake leptons can originate from decays of bottom or
charm hadrons, jets misidentified as electrons, leptons from kaon
or pion decays, or electrons from photon conversions.
 
All background contributions are estimated by using MC
samples that are normalized to their respective SM predicted cross-sections
calculated at NLO in QCD.
The cross-section of the \ttH background includes
NLO+NLL soft-gluon resummation~\cite{Kulesza:2015vda}.
For the $\ttbar+\tW$ nonprompt lepton backgrounds the
normalization is extracted from data, as described later.
 
After applying the event selection requirements, diboson, \ttZ and \SMtZq
production constitute the largest backgrounds.
For SR1, the dominant backgrounds are \ttZ and \VVHF production.
Monte Carlo simulation indicates that these represent more than \SI{65}{\%} of the
total number of selected background events in this region, with the
two processes contributing equally.
For SR2, \VVHF and \SMtZq are the dominant
backgrounds, giving \SI{70}{\%} of background events.
The processes with nonprompt leptons constitute a minor background,
with their contribution being at most \SI{10}{\%} of the total
selected events.

Four control regions (CRs) are defined and used in the fit that is
described in \cref{sec:results}. The CRs are used to adjust the
normalization and to reduce the associated systematic uncertainties in
the main backgrounds. The selections applied to define the CRs are
summarized in \cref{tab:cr:cutsummary} and described in the
following.
 
A \ttbar CR is designed to control the \ttbar background.
The \ttbar CR is constructed by requiring the presence of three
leptons, with one of the possible pairs having opposite charge, as in the SRs. To veto the
presence of a \PZ boson, 
the opposite-sign lepton pair is also required to consist of different flavors.
Events with at least one jet, with exactly one
\Pqb-tagged, are considered. This region is dominated by \ttbar events with
\SI{40}{\%} contamination from other backgrounds, mainly \ttW and \ttH.
A total of 157 data events are selected for the \ttbar CR.
 
To control the \ttZ background, a \ttZ CR is defined. The
requirements on the leptons are the same as for the SRs, while at
least four jets, with exactly two \Pqb-tagged, are required.
This region is dominated by \ttZ events with \SI{25}{\%} contamination
from other backgrounds, mainly \SMtZq and \VVHF.
A total of 286 data events are selected for the \ttZ CR.
 
Two mass sideband CRs are also included. These CRs are designed to
contain a mixture of the main background sources (\ttZ and diboson).
The mass sideband CR1 is defined with almost the same event selection as
SR1, with the differences being that the mass of the FCNC top-quark
candidate must be more than 2\sigmaFCNC from
\mtopvalue, and the mass of the SM top-quark candidate must
also be more than 2\sigmaSM from \mtopvalue.
The mass sideband CR2 is defined with almost the same event selection as
SR2, with the differences being that only events with one jet are
considered and that the mass of the SM top-quark candidate must be
more than 2\sigmaSM from \mtopvalue.
Totals of 343 and 104 data events are selected for the mass
sidebands CR1 and CR2 respectively.
 
\begin{table}[htbp]
\caption{
Overview of the requirements applied to select the events in the control regions.
OSSF is an opposite-sign same-flavor lepton pair,
$m_{\PZ}$ = \SI{91.2}{\GeV} and $m_{\Pqt}$ = \mtopvalue.
}
\label{tab:cr:cutsummary}
\centering
\small
\begin{tabular}{cccc}
\toprule
\multicolumn{4}{c}{Common selections} \\
\midrule
\multicolumn{4}{c}{Exactly 3 leptons with $\pT(\ell_1)> \SI{27}{\GeV}$} \\
\toprule
\ttbar CR & \ttZ CR & Sideband CR1 & Sideband CR2 \\
\midrule
$\ge$ 1 \OS pair, no \OSSF  & $\ge$ 1 \OSSF pair & $\ge$ 1 \OSSF pair & $\ge$ 1 \OSSF pair \\
& with $|\mll - m_{\PZ}| < \SI{15}{\GeV}$ & with $|\mll - m_{\PZ}| < \SI{15}{\GeV}$ & with $|\mll - m_{\PZ}| < \SI{15}{\GeV}$ \\
-- & -- & --  & $\mT(\ell_{\PW},\Pgn) > \SI{40}{\GeV}$ \\
$\ge$ 1 jet & $\ge$ 4 jets & $\ge$ 2 jets & 1 jet \\
1 \Pqb-jet  & 2 \Pqb-jets  & 1 \Pqb-jet  & 1 \Pqb-jet \\
-- & -- & $|m^{\mathrm{reco}}_{j_a\ell\ell} - m_{\Pqt}| > 2\sigmaFCNC$ & -- \\
-- & -- & $|m^{\mathrm{reco}}_{j_b\ell_W\nu} - m_{\Pqt}| > 2\sigmaSM$ & $|m^{\mathrm{reco}}_{j_b\ell_W\nu} - m_{\Pqt}| > 2\sigmaSM$ \\
\bottomrule
\end{tabular}
\end{table}

To better separate the signal from the backgrounds, a multivariate
analysis (MVA) technique is used. The chosen MVA is the gradient
boosted decision tree (GBDT) method implemented with TMVA~\cite{GBDT,TMVA}.
Decision trees~\cite{DT} recursively partition the parameter space
into regions where signal or background purities are
enhanced. Gradient boosting is a method which improves the performance and
stability of decision trees and involves the combination of many trees
into a single final discriminant. After boosting, the final score
undergoes a transformation to map the scores onto the interval $-1$ to $+1$.
The most signal-like events have scores near $+1$ while the most
background-like events have scores near $-1$.
A $k$-fold cross validation is employed.
 
The GBDT training is done separately for the LH and RH samples
and in each SR as follows.
In SR1, for both the FCNC \tZu and \tZc coupling searches, the expected
contribution from FCNC processes in \ttbar decay is
significantly higher than the one from single-top-quark
production. Therefore, the GBDT is trained with only the
FCNC-in-\ttbar-decay signal against all backgrounds. Since the
kinematics of FCNC-in-\ttbar-decay events for \tZu and \tZc couplings
are similar, the FCNC-in-\ttbar-decay signal samples with the two couplings are combined to train the GBDT.
Therefore, in SR1 a single MVA discriminant, \Done, is built for both
the FCNC \tZu and \tZc coupling searches.
In contrast, SR2 is particularly sensitive to the FCNC \tZu coupling in
single-top-production events. Thus, the corresponding MVA
discriminant, \DtwoU, is built by training the GBDT with the
the \tZu-coupling FCNC-in-single-top-production sample against
all backgrounds.
Despite the lower sensitivity to the FCNC \tZc coupling in SR2, this
region is used in combination with SR1 in the search for a FCNC \tZc coupling signal.
In the total expected FCNC \tZc signal yield, the contribution from
the FCNC processes in \ttbar decay events is comparable to the one from
the single-top-quark-production events. Therefore, in SR2
the MVA discriminant for the search for a FCNC \tZc coupling signal, \DtwoC, is built using
both the FCNC-in-\ttbar-decay and FCNC-in-single-top-production samples
against all backgrounds.
 
For the training of each of the three discriminants, a total of six variables
is used. 
Only variables
that provide good separation and are well modeled are used in the final training.
For the \Done discriminant the six variables are: the reconstructed masses of the SM
and FCNC top-quark candidates, the $\Delta R$ separation between them,
the $\Delta R$ separation between the lepton from the SM top-quark
decay and the reconstructed \PZ boson, the number of jets, and the
transverse momentum of the jets associated with the \Pqu/\Pqc-quark from
the FCNC top-quark candidate's decay.
For both the \DtwoU and \DtwoC discriminants the following six variables are used: the \pT of
the \PZ boson and of the \Pqb-tagged jet, the $\Delta R$ separation
between them, the SM top-quark candidate's mass, the $\Delta R$
separation between the lepton from the SM top-quark candidate decay
and the reconstructed \PZ boson, and the $\chi^2$ from the kinematic fit under the
signal hypothesis of an FCNC process in single-top-quark production.

In order to check the stability of the fit that will be presented in
\cref{sec:results}, two validation regions (VRs) are defined.
The events selected by the VRs are a subset of the SRs previously
described and they are obtained by applying a selection on the GBDT discriminants.
VR1 is defined by selecting events with $\Done < -0.6$ from the SR1, while
VR2 contains events from SR2 with $\DtwoU < -0.7$ and $\DtwoC < -0.4$.
With the given normalization of the signal samples,
the fraction of signal events that is selected from the SRs to enter the VRs
ranges from \SI{2}{\%} to \SI{5}{\%}, depending on the SR.
The signal contamination in the VRs is at most \SI{2}{\%}.


\FloatBarrier

 
\section{Systematic uncertainties}
\label{sec:sys}
 
Systematic uncertainties in the signal acceptance
and in the normalization of the individual backgrounds,
as well as uncertainties in the shape of the fitted distributions,
are taken into account. These are treated as being correlated among
the different regions, unless stated otherwise.
The uncertainties are classified into the following categories:
 
\paragraph{Reconstruction efficiency and calibration uncertainties:}
Systematic uncertainties affecting the reconstruction efficiency and
energy calibration of electrons, muons, jets and \Pqb-jets are propagated through the analysis.
 
The differences between the electron (muon) trigger, reconstruction,
selection and isolation efficiencies in data and those in MC simulation
are corrected for by scale factors derived from dedicated
\( \PZ\rightarrow \ee \) (\(~\PZ~\rightarrow~\mumu~\))
enriched control samples using a tag-and-probe method~\cite{EGAM-2018-01,MUON-2018-03-INT}.
Uncertainties in these scale factors are taken into account.
Moreover, uncertainties are included for the electron (muon)
energy (momentum) scale and resolution~\cite{EGAM-2018-01,MUON-2018-03-INT}.
 
For the jets, an uncertainty for the JVT requirement is considered.
The jet energy scale was derived using information from
test-beam data, LHC collision data and simulation, as described in Ref.~\cite{PERF-2016-04}.
The impact of the uncertainty in the jet energy resolution is
also evaluated.
 
The \btag efficiencies and mistagging rates are measured in data
using the same methods as described in
Refs.~\cite{FTAG-2018-01,ATLAS-CONF-2018-001,ATLAS-CONF-2018-006},
with the systematic uncertainties due to \btag efficiency and the mistagging rates calculated separately.
The impact of the uncertainties on the \btag calibration is evaluated
separately for \Pqb-, \Pqc- and light-jets in the MC samples.
 
The uncertainty in \MET due to a possible miscalibration
of the soft-track component of the \MET is derived from data--MC comparisons
of the \pT balance between the hard and soft \MET components~\cite{PERF-2016-07}.
The uncertainty associated with the leptons and jets is propagated from the
corresponding uncertainties in the energy/momentum scales and resolutions, and
is classified together with the uncertainty associated with the corresponding objects.
 
\paragraph{Signal and background modeling:}
The systematic uncertainties due to MC modeling of the signal and the
main backgrounds are estimated by comparing samples from different MC generators
and PDF sets
and by varying the parameters associated with the renormalization and
factorization scales, and additional radiation.
For some processes, some of these uncertainties are found to be
negligible and therefore they are not mentioned in the following.
 
For the signal, the effects of the systematic
uncertainty in the renormalization and factorization scales,
\muR and \muF, are taken into account by varying these parameters by
factors of \SI{2}{} and \SI{0.5}{} with respect to their default values
and comparing the results of these variations with the nominal prediction.
The uncertainty in the modeling of the parton shower is estimated
by comparing the nominal signal sample with one generated with
\HERWIG[7] instead of \PYTHIA[8].
PDF uncertainties are found to be negligible and are not included for
the signal.
 
For the \ttZ and \SMtZq backgrounds, the following uncertainties are included.
The effect of changing the parton shower is considered as an
uncertainty, following the same strategy used for the signal.
The uncertainty due to ISR is estimated by
comparing the nominal event sample with two samples where the Var3c
up and down variations of the A14 tune were employed.
Uncertainties from the variation of \muR and \muF are also included.
 
For the \tWZ background, the effect of changing the modeling of the
interference with \ttZ is included by comparing two different
diagram removal predictions.
 
The effect of changing the MC generator for the modeling of the
diboson background is considered as an uncertainty.
It is evaluated by comparing the nominal
\SHERPA sample with one generated with \POWHEGBOX.
This uncertainty is split into the two light- and heavy-flavor
components and evaluated separately for each jet multiplicity.
Uncertainties in the \muR and \muF scales, as well as in the PDF and
in \alphas are also included for the diboson background.
 
For the \ttbar background, several sources of uncertainty are taken
into account.
The effect of changing the parton shower is included as an
uncertainty.
The Var3c A14 tune variations, as well as variations of \muR and \muF
are also included.
Additionally, the uncertainty associated with the \hdamp\ parameter is
evaluated by using the alternative sample with the \hdamp\ value
increased to $3\, \mtop$.
The \NNPDF[3.0lo] replicas are used to evaluate the PDF uncertainties
for the nominal PDF.
Finally, an uncertainty is added to take into account the
differences in \ttbar background composition between the SRs
and the \ttbar CR, which is used to control the \ttbar background in the fit to data.
In particular, the fractions of nonprompt leptons originating from each source
are computed, separately for photon conversions and \Pqb-hadron
decays, in the SRs and in the \ttbar CR, for each jet
multiplicity.
Then the maximum variation of the fractions between the control region
and the signal regions is taken as an uncertainty.

\paragraph{Signal and background rate uncertainty:}
The \ttbar cross-section uncertainties due to the PDF and
\alphas are calculated using the \PDFforLHC[15] prescription~\cite{Butterworth:2015oua}
with the \MSTW[nnlo]~\cite{Martin:2009iq,Martin:2009bu},
\CT[10nnlo]~\cite{Lai:2010vv,Gao:2013xoa} and
\NNPDF[2.3lo] PDF sets,
and are added in quadrature to the effect of the scale uncertainty, resulting in a total
uncertainty of \SI{5.5}{\%} that is assigned to the FCNC-in-\ttbar-decay signal.
 
For the \ttZ background, a \SI{12}{\%} rate uncertainty is
included~\cite{LHCHiggsCrossSectionWorkingGroup:2016ypw},
and for the \ttH process the normalization uncertainty is
\SI{15}{\%}~\cite{LHCHiggsCrossSectionWorkingGroup:2016ypw},
while for \ttW a more conservative \SI{50}{\%} is
used~\cite{TOPQ-2018-05}. 
For the \SMtZq process, an uncertainty of \SI{15}{\%} in the
normalization is applied~\cite{TOPQ-2018-01,CMS-TOP-16-020}, while for
the \tWZ process a more conservative \SI{30}{\%} is used.
For \VVLF production, the normalization uncertainty is taken to be
\SI{20}{\%}~\cite{STDM-2018-03} and for \VVHF production it is
\SI{30}{\%}~\cite{HIGG-2016-29}.
Concerning the \Zjets process, a rate uncertainty of \SI{100}{\%}
is applied, due to the presence of a nonprompt lepton.
A conservative overall normalization uncertainty of \SI{50}{\%} is applied
to the remaining minor backgrounds (\threetops, \fourtops, \VVV, \VH and \ttWW).
These background components are typically well below \SI{1}{\%} in the SRs.
 
\paragraph{Luminosity:}
The uncertainty in the combined 2015--2018 integrated luminosity
is \SI{1.7}{\%}~\cite{ATLAS-CONF-2019-021},
obtained using the LUCID-2 detector~\cite{LUCID2} for the primary luminosity measurements.
 
\paragraph{Uncertainty in \pileup modeling:}
The uncertainty in \pileup modeling is accounted for by varying the reweighting of the MC samples to the data \pileup conditions,
using the uncertainty in the average number of interactions per bunch crossing.


 
\section{Results}
\label{sec:results}
 
A simultaneous binned profile likelihood fit to the data in the SRs
and the CRs is performed using MC distributions of both the signal and
background predictions.
Four separate fits are performed to extract LH and RH results for the
FCNC \tZu and \tZc couplings.
Only the relevant signal templates are used in each fit.
In the fit to extract limits on the FCNC \tZu coupling,
the templates are binned distributions of the \Done discriminant in SR1 and
in the mass sideband CR1; the \DtwoU discriminant in SR2 and in the
mass sideband CR2; and the total event yields in the \ttbar CR and the
\ttZ CR. When fitting to extract limits on the FCNC \tZc coupling, the
\DtwoC discriminant is used instead of \DtwoU.
 
The fitted SRs are defined from the SRs
described in \cref{sec:evtsel} after removing events that
constitute VRs described in \cref{sec:mva} that are not included in the fit,
but the fit results are propagated to those regions.
The signal selection efficiency for the FCNC-in-\ttbar-decay signal in
SR1 ranges between \SI{4}{\%} and \SI{5}{\%}, while that for the
FCNC-in-single-top-production signal in SR2 ranges between \SI{3}{\%}
and \SI{4}{\%}. In contrast, the signal selection efficiency for the
FCNC-in-\ttbar-decay signal in SR2 and the
FCNC-in-single-top-production signal in SR1 is around \SI{1}{\%}.
 
The statistical analysis used to extract the signal is based on a binned
likelihood function $\mathcal{L}(\mu,\vec{\theta})$ constructed as a product of
Poisson probability terms over all bins in each considered
distribution, and Gaussian constraint terms for $\vec{\theta}$, a set of
nuisance parameters that parameterize effects of MC statistical and
systematic uncertainties in the signal and background expectations.
The signal strength parameter $\mu$ is a multiplicative factor applied to the
number of signal events normalized to a reference branching ratio.
For that, the most stringent limits mentioned in \cref{sec:intro} are used.
The nuisance parameters are allowed to vary in the combined fit to adjust the
expectations for signal and background according to the corresponding
systematic uncertainties, and their final values are the adjustment
that best fits the data.
The normalization of the $\ttbar + \tW$ backgrounds is unconstrained in the fit.
 
A test statistic, $\tilde{q}_\mu$, is constructed according to the profile
likelihood ratio:
\begin{equation}
\tilde{q}_{\mu} = \left\{ \begin{array}{ll} -2 \ln \left(  \frac{\mathcal{L}\left(\mu, \hat{\hat{\vec{\theta}}}(\mu)\right)}{\mathcal{L}\left(0, \hat{\hat{\vec{\theta}}}(0)\right)}\right) &
\mathrm{if}\ \hat{\mu} < 0, \\
-2 \ln \left(  \frac{\mathcal{L}\left(\mu, \hat{\hat{\vec{\theta}}}(\mu)\right)}{\mathcal{L}\left(\hat{\mu}, \hat{\vec{\theta}}\right)}\right) & \mathrm{if}\ 0 \leq \hat{\mu} \leq \mu, \\
\ \ 0 & \mathrm{if}\ \hat{\mu}>\mu, \end{array}  \right.
\label{eq:q_mu}
\end{equation}
where $\hat{\mu}$ and $\hat{\vec{\theta}}$ are the parameters that
maximize the likelihood,
and $\hat{\hat{\vec{\theta}}}$ are the nuisance parameter values that
maximize the likelihood for a given $\mu$ hypothesis.
This test statistic is used to determine the probability for accepting
the background-only hypothesis for the observed data.
 
\Cref{tab:results:tzu:lh:yields:vrs} shows the pre- and post-fit
predictions for the signal and background event yields along with the
observed numbers of events in the VRs.
The post-fit yields refer to the fit for the FCNC \tZu LH coupling extraction.
The data and background expectation are in better agreement after the
fit, with an increase of the \VVHF background normalization within its pre-fit uncertainty.
The post-fit level of agreement between data and the background
prediction in the VRs shows no significant mismodeling.
 
\begin{table}[htbp]
\caption{Predicted and observed yields in the two VRs
considered in the fit.
The signal and background predictions are shown before
(\enquote{Pre-fit}) and after (\enquote{Post-fit}) the fit to
data for the FCNC \tZu LH coupling extraction.
The quoted uncertainties include the statistical and systematic
uncertainties of the yields. For the post-fit predictions, they are computed taking into account
correlations among nuisance parameters and among processes.
For the backgrounds with a nonprompt or fake lepton, the contribution from
\ttbar + \tW is shown separately from \enquote{Other fakes}.
For the minor backgrounds, the contribution from \ttW and \ttH are
shown separately from \enquote{Other bkg.}.
}
\label{tab:results:tzu:lh:yields:vrs}
\centering
\footnotesize
\begin{tabular}{l
S[table-format=3.2]@{$\,\pm\,$}S[table-format=2.2]
S[table-format=2.3]@{$\,\pm\,$}S[table-format=2.3]
S[table-format=3.2]@{$\,\pm\,$}S[table-format=2.2]
S[table-format=2.3]@{$\,\pm\,$}S[table-format=1.3]
}
\toprule
& \multicolumn{4}{c}{Pre-fit} & \multicolumn{4}{c}{Post-fit} \\
& \multicolumn{2}{c}{VR1} & \multicolumn{2}{c}{VR2} &\multicolumn{2}{c}{VR1} &\multicolumn{2}{c}{VR2} \\
\midrule
\ttZ + \tWZ   & 70 & 10 & 2.2 & 0.6 & 70 & 7 & 2.4 & 0.6 \\
\VVLF   & 10 & 5 & 9.8 & 3.4 & 10 & 5 & 9.7 & 3.0 \\
\VVHF   & 56 & 28 & 36 & 14 & 60 & 14 & 47 & 8 \\
\SMtZq   & 6.5 & 1.6 & 13.5 & 2.7 & 6.6 & 1.5 & 14.7 & 2.6 \\
\ttbar + \tW fakes  & 5.4 & 2.6 & 4.5 & 1.7 & 4.8 & 2.1 & 3.8 & 1.4 \\
Other fakes   & 0.0 & 0.6 & 1.4 & 1.9 & 0.03 & 0.24 & 0.8 & 1.1 \\
\ttW   & 2.3 & 1.2 & 0.48 & 0.26 & 2.3 & 1.2 & 0.48 & 0.25 \\
\ttH   & 3.0 & 0.5 & 0.101 & 0.032 & 3.0 & 0.5 & 0.108 & 0.033 \\
Other  bkg. & 0.8 & 0.4 & 0.5 & 0.7 & 0.8 & 0.4 & 0.5 & 0.6 \\
\midrule
Total background  & 154 & 31 & 69 & 15 & 158 & 13 & 79 & 7 \\
\midrule
Data   & \multicolumn{2}{c}{151} & \multicolumn{2}{c}{80} & \multicolumn{2}{c}{151} & \multicolumn{2}{c}{80} \\
\midrule
Data / Bkg.   & 0.98 & 0.22 & 1.16 & 0.29 & 0.96 & 0.11 & 1.01 & 0.15  \\
\bottomrule
\end{tabular}
\end{table}
 
\Cref{tab:results:tzu:lh:yields:srs:postfit,tab:results:tzu:lh:yields:crs:postfit}
show the observed number of
events in data and the post-fit predictions for the signal and
background event yields in the SRs and CRs.
The yields refer to the fit for the FCNC \tZu LH coupling extraction.
Good agreement between data and the SM expectation is observed.
The normalization factor for the $\ttbar + \tW$
backgrounds, which is an unconstrained
fit parameter, agrees with unity within uncertainties.
The variations of the post-fit background normalizations are within
pre-fit uncertainties.
All post-fit values of the nuisance parameters are less than one standard
deviation from the pre-fit values.
The statistical component is the dominant contribution in the total
uncertainty.
The same conclusions are obtained from the fits for the other FCNC
couplings.

\begin{table}[htbp]
\caption{Predicted and observed yields in the two SRs
considered in the fit.
The signal and background predictions are shown after the fit to
data for the FCNC \tZu LH coupling extraction.
The quoted uncertainties include the statistical and systematic
uncertainties of the yields, computed taking into account
correlations among nuisance parameters and among processes.
For the backgrounds with a nonprompt or fake lepton, the contribution from
$\ttbar + \tW$ is shown separately from \enquote{Other fakes}.
For the minor backgrounds, the contribution from \ttW and \ttH are
shown separately from \enquote{Other bkg.}.
}
\label{tab:results:tzu:lh:yields:srs:postfit}
\centering
\begin{tabular}{l
S[table-format=3.1]@{$\,\pm\,$}S[table-format=2.1]
SSSSS[table-format=3.2]@{$\,\pm\,$}S[table-format=2.2]
}
\toprule
& \multicolumn{2}{c}{SR1} & & & & \multicolumn{3}{c}{SR2} \\
& \multicolumn{2}{c}{$(\Done > -0.6)$} & \multicolumn{6}{r}{$(\DtwoU > -0.7$ or $\DtwoC > -0.4)$} \\
\midrule
\ttZ + \tWZ   & 137 & 12 & & & & & 36 & 6 \\
\VVLF   & 18 & 7 & & & & & 24 & 8 \\
\VVHF   & 114 & 19 & & & & & 162 & 26 \\
\SMtZq   & 46 & 7 & & & & & 108 & 18 \\
\ttbar + \tW fakes  & 14 & 4 & & & & & 27 & 8 \\
Other fakes   & 7 & 8 & & & & & 5 & 6 \\
\ttW   & 4.2 & 2.1 & & & & & 3.1 & 1.6 \\
\ttH   & 4.8 & 0.7 & & & & & 0.89 & 0.17 \\
Other bkg.  & 2.0 & 1.0 & & & & & 2.5 & 2.9 \\
\midrule
FCNC $(u)tZ$   & 0.9 & 1.7 & & & & & 4 & 8 \\
FCNC $\ttbar(uZ)$   & 5 & 9 & & & & & 0.8 & 1.5 \\
\midrule
Total background  & 348 & 15 & & & & & 369 & 21 \\
\midrule
Data   & \multicolumn{2}{c}{345} & & & & \multicolumn{3}{c}{380} \\
\bottomrule
\end{tabular}
\end{table}
 
\begin{table}[htbp]
\caption{Predicted and observed yields in the four CRs
considered in the fit.
The signal and background predictions are shown after the fit to
data for the FCNC \tZu LH coupling extraction.
The quoted uncertainties include the statistical and systematic
uncertainties of the yields, computed taking into account
correlations among nuisance parameters and among processes.
For the backgrounds with a nonprompt or fake lepton, the contribution from
$\ttbar + \tW$ is shown separately from \enquote{Other fakes}.
For the minor backgrounds, the contribution from \ttW and \ttH are
shown separately from \enquote{Other bkg.}.
}
\label{tab:results:tzu:lh:yields:crs:postfit}
\centering
\footnotesize
\begin{tabular}{l
S[table-format=3.2]@{$\,\pm\,$}S[table-format=2.2]
S[table-format=3.2]@{$\,\pm\,$}S[table-format=1.2]
S[table-format=3.2]@{$\,\pm\,$}S[table-format=2.2]
S[table-format=3.3]@{$\,\pm\,$}S[table-format=2.3]
}
\toprule
& \multicolumn{2}{c}{Sideband CR1} & \multicolumn{2}{c}{Sideband CR2} & \multicolumn{2}{c}{\ttZ CR} & \multicolumn{2}{c}{\ttbar CR}\\
\midrule
\ttZ + \tWZ   & 102 & 14 & 8.2 & 1.4 & 230 & 18 & 15.4 & 1.5 \\
\VVLF   & 27 & 11 & 12 & 4 & 0.23 & 0.19 & 0.38 & 0.25 \\
\VVHF   & 166 & 25 & 64 & 9 & 17 & 8 & 2.9 & 0.5 \\
\SMtZq   & 22 & 4 & 6.8 & 1.4 & 21 & 5 & 0.96 & 0.19 \\
\ttbar + \tW fakes  & 9.3 & 2.6 & 7.2 & 2.1 & 4.0 & 1.3 & 93 & 19 \\
Other fakes   & 2 & 4 & 2.0 & 2.8 & 0.15 & 0.18 & 0.08 & 0.09 \\
\ttW   & 4.5 & 2.3 & 2.3 & 1.2 & 3.0 & 1.5 & 27 & 13 \\
\ttH   & 2.6 & 0.4 & 0.33 & 0.07 & 7.5 & 1.2 & 14.1 & 2.2 \\
Other bkg.  & 3.3 & 2.5 & 0.8 & 0.4 & 1.9 & 0.9 & 3.2 & 1.5 \\
\midrule
FCNC $(u)tZ$   & 0.4 & 0.7 & 0.17 & 0.33 & 0.09 & 0.18 & 0.05 & 0.10 \\
FCNC $\ttbar(uZ)$   & 0.14 & 0.27 & 0.04 & 0.07 & 0.11 & 0.20 & 0.018 & 0.035 \\
\midrule
Total background  & 338 & 18 & 104 & 8 & 284 & 16 & 157 & 13 \\
\midrule
Data   & \multicolumn{2}{c}{343} & \multicolumn{2}{c}{104} & \multicolumn{2}{c}{286} & \multicolumn{2}{c}{157} \\
\bottomrule
\end{tabular}
\end{table}
 
The $\mu$ parameters are shown in \cref{tab:results:mu}.
 
\begin{table}[htbp]
\caption{Summary of the signal strength $\mu$ parameters obtained from the fits to
extract LH and RH results for the FCNC \tZu and \tZc couplings.
For the reference branching ratio, the most stringent limits are
used~\cite{TOPQ-2017-06}.
}
\label{tab:results:mu}
\centering
\begin{tabular}{ccc}
\toprule
Vertex & Coupling & $\mu$ \\
\midrule
\tZu & LH & $0.08\pm0.12~\mathrm{(stat.)}\pm0.08~\mathrm{(syst.)}$ \\
\tZu & RH & $0.10\pm0.12~\mathrm{(stat.)}\pm0.08~\mathrm{(syst.)}$ \\
\tZc & LH & $0.10\pm0.17~\mathrm{(stat.)}\pm0.14~\mathrm{(syst.)}$ \\
\tZc & RH & $0.06\pm0.16~\mathrm{(stat.)}\pm0.13~\mathrm{(syst.)}$ \\
\bottomrule
\end{tabular}
\end{table}
 
\begin{figure}[htbp]
\centering
\begin{subfigure}[b]{.45\textwidth}
\includegraphics[width=1\textwidth]{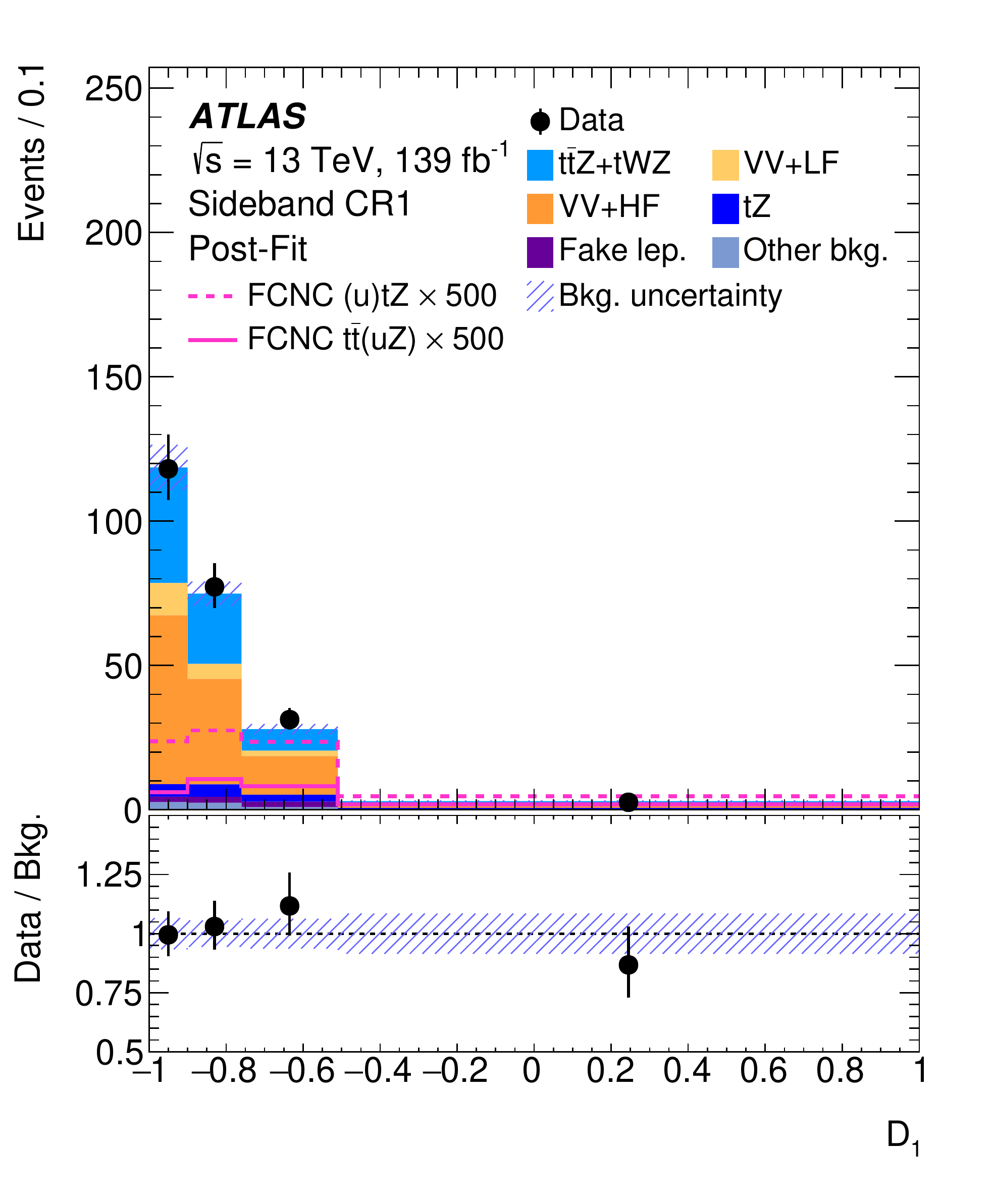}
\label{subfig:results:tzu:lh:crplots-1}
\caption{ }
\end{subfigure}
\begin{subfigure}[b]{.45\textwidth}
\includegraphics[width=1\textwidth]{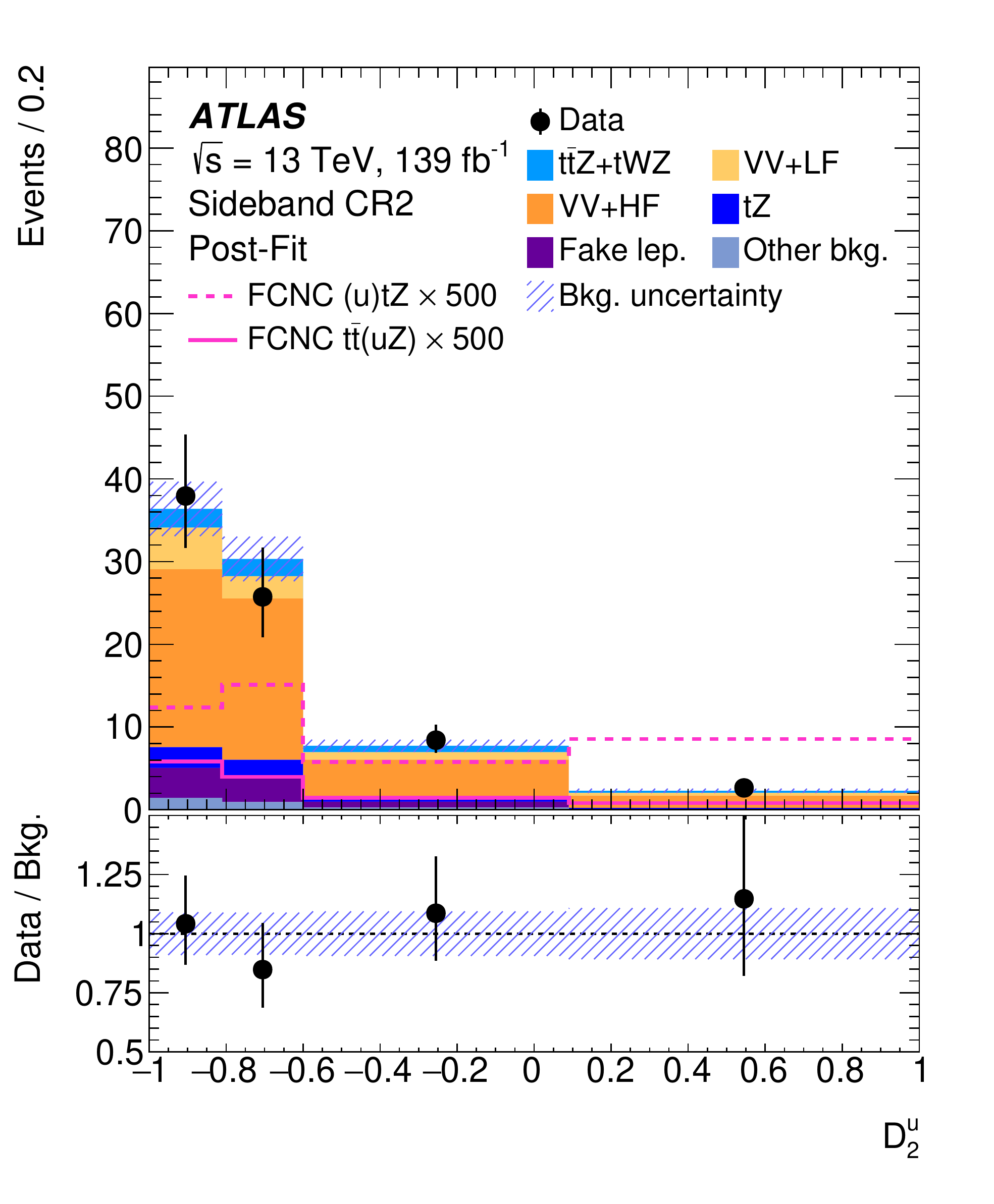}
\label{subfig:results:tzu:lh:crplots-2}
\caption{ }
\end{subfigure}
\begin{subfigure}[b]{.45\textwidth}
\includegraphics[width=1\textwidth]{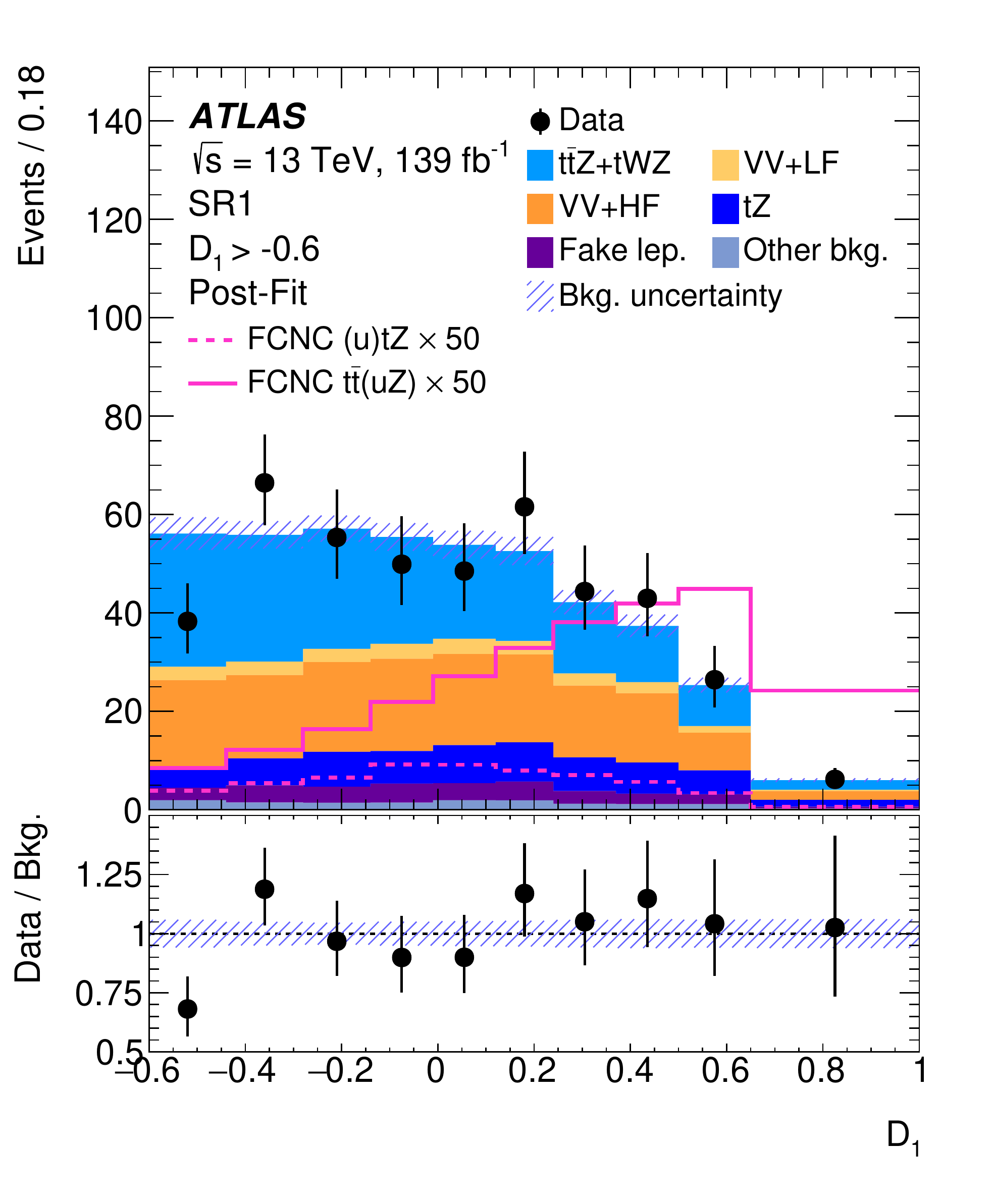}
\label{subfig:results:tzu:lh:srplots-1}
\caption{ }
\end{subfigure}
\begin{subfigure}[b]{.45\textwidth}
\includegraphics[width=1\textwidth]{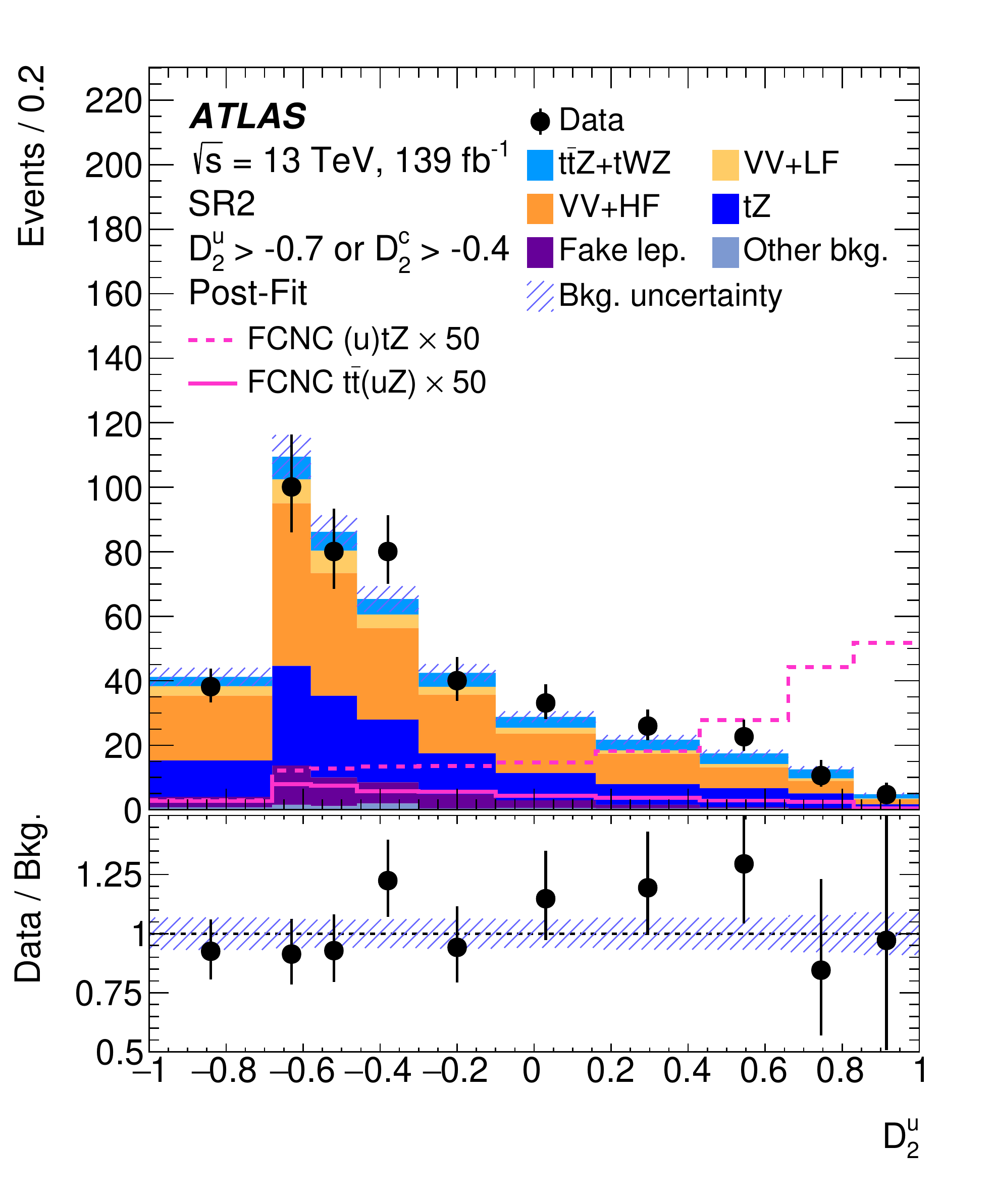}
\label{subfig:results:tzu:lh:srplots-2}
\caption{ }
\end{subfigure}
\caption{Comparison between data and background prediction
after the fit to data (\enquote{Post-Fit})
for the FCNC \tZu LH coupling extraction
for the fitted distributions in the CRs and SRs.
The distributions are:
(a) the \Done discriminant in the mass sideband CR1,
(b) the \DtwoU discriminant in the mass sideband CR2,
(c) the \Done discriminant in SR1 and
(d) the \DtwoU discriminant in SR2.
The uncertainty band includes both the statistical and systematic
uncertainties in the background prediction.
The FCNC \tZu LH signals are also separately shown, normalized to
500 or 50 times the best fit of the signal yield.
The lower panels show the ratios of the data (\enquote{Data})
to the background prediction (\enquote{Bkg.}).
}
\label{fig:results:tzu:lh:plots}
\end{figure}

\begin{figure}[htbp]
\centering
\begin{subfigure}[b]{.45\textwidth}
\includegraphics[width=1\textwidth]{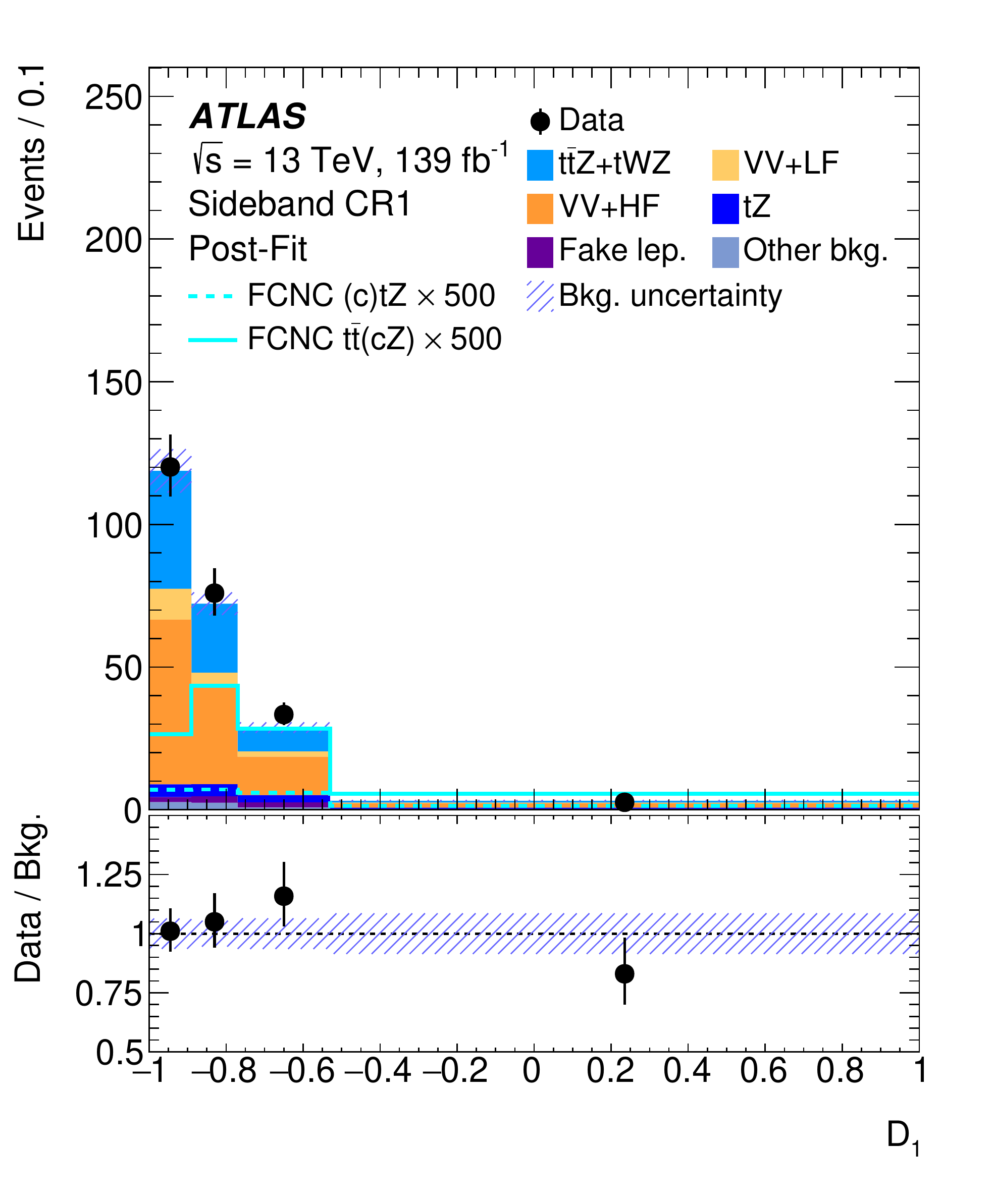}
\label{subfig:results:tzc:lh:crplots-1}
\caption{ }
\end{subfigure}
\begin{subfigure}[b]{.45\textwidth}
\includegraphics[width=1\textwidth]{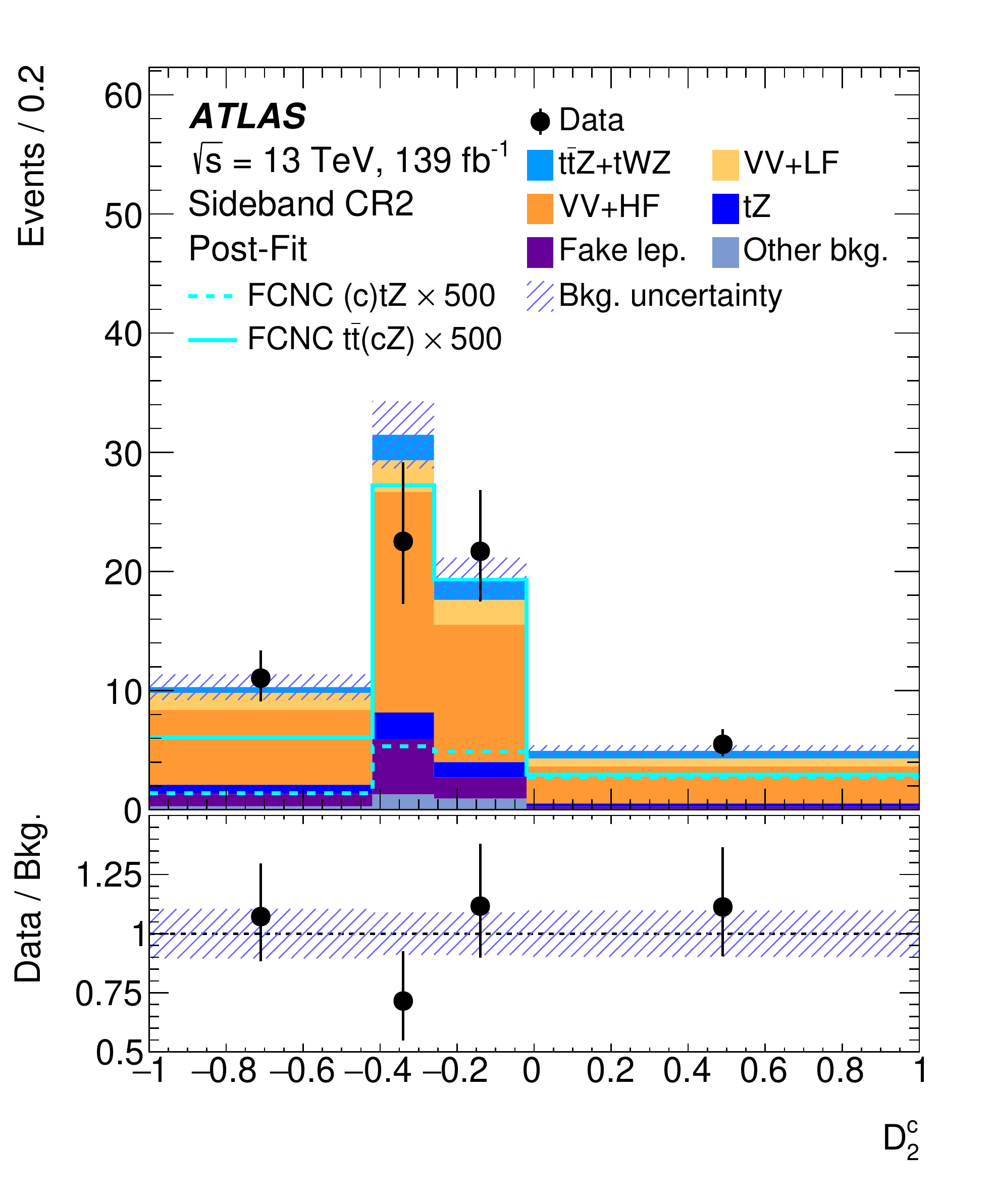}
\label{subfig:results:tzc:lh:crplots-2}
\caption{ }
\end{subfigure}
\begin{subfigure}[b]{.45\textwidth}
\includegraphics[width=1\textwidth]{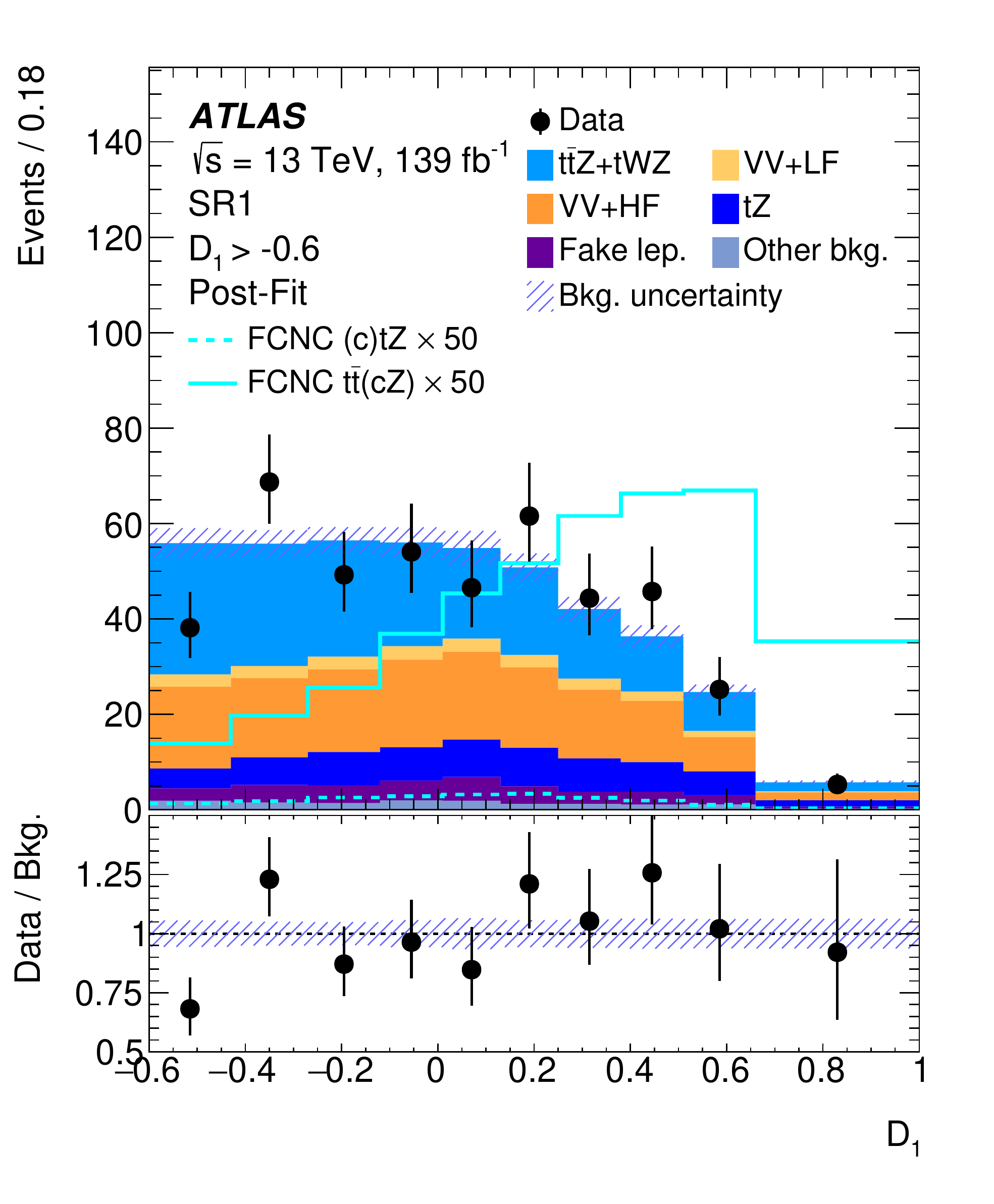}
\label{subfig:results:tzc:lh:srplots-1}
\caption{ }
\end{subfigure}
\begin{subfigure}[b]{.45\textwidth}
\includegraphics[width=1\textwidth]{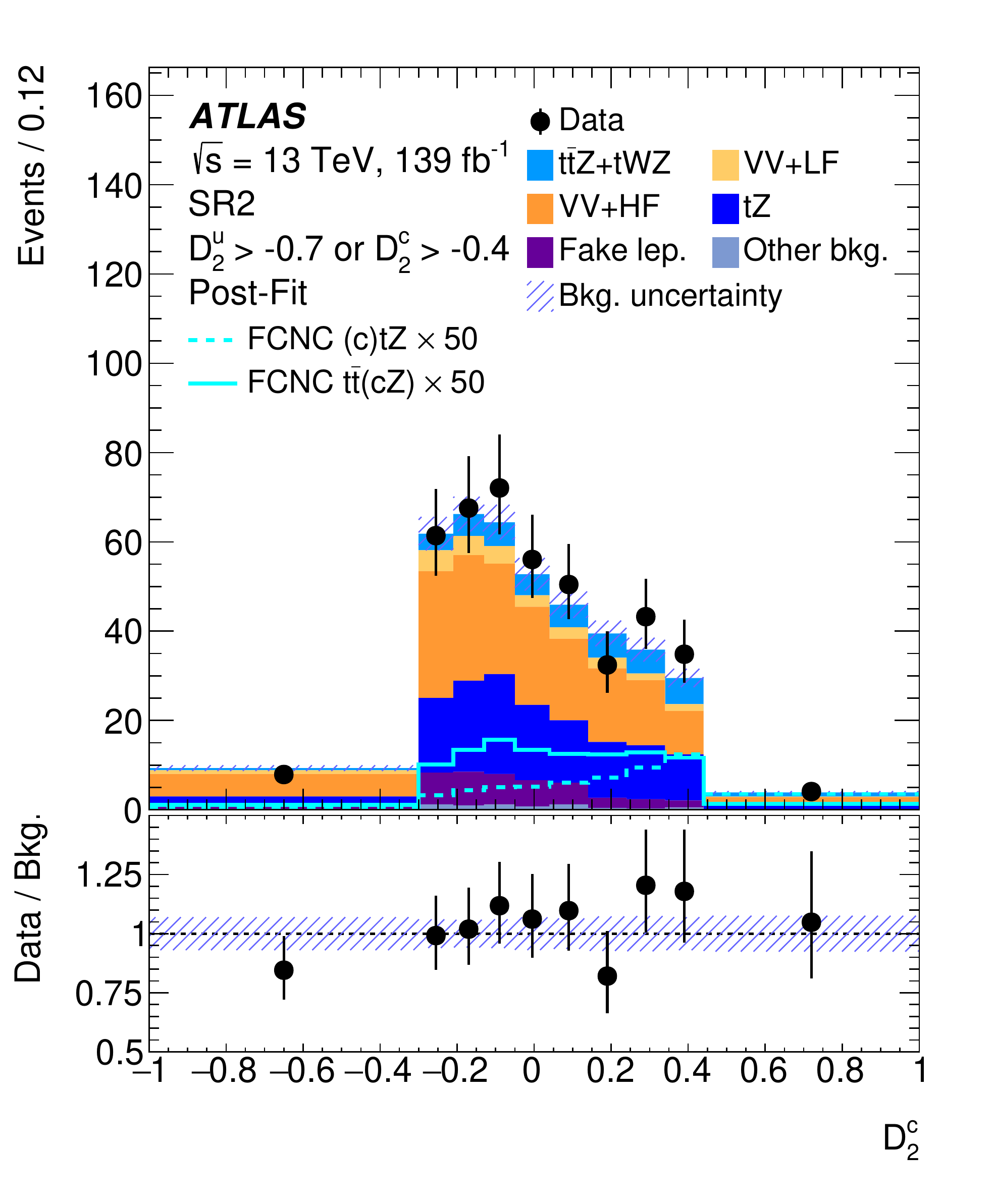}
\label{subfig:results:tzc:lh:srplots-2}
\caption{ }
\end{subfigure}
\caption{Comparison between data and background prediction
after the fit to data (\enquote{Post-Fit})
for the FCNC \tZc LH coupling extraction
for the fitted distributions in the CRs and SRs.
The distributions are:
(a) the \Done discriminant in the mass sideband CR1,
(b) the \DtwoC discriminant in the mass sideband CR2,
(c) the \Done discriminant in SR1 and
(d) the \DtwoC discriminant in SR2.
The uncertainty band includes both the statistical and systematic
uncertainties in the background prediction.
The FCNC \tZc LH signals are also separately shown, normalized to
500 or 50 times the best fit of the signal yield.
The lower panels show the ratios of the data (\enquote{Data})
to the background prediction (\enquote{Bkg.}).
}
\label{fig:results:tzc:lh:plots}
\end{figure}

\Cref{fig:results:tzu:lh:plots}
shows the distributions of the fitted variables in the CRs and
SRs after the fit for the FCNC \tZu LH coupling extraction.
For the FCNC \tZc LH coupling extraction, the fitted distributions are
presented in \Cref{fig:results:tzc:lh:plots},
where \DtwoC is used in SR2 and in the mass sideband CR2.
For the \ttbar and \ttZ CRs, only the event yields are used.
The data and background
prediction agree within the uncertainties.
 
Limits on each FCNC \ttoZq branching ratio are computed with the
CL$_\text{s}$ method \cite{Read:2002hq} using the asymptotic
properties of $q_\mu$ \cite{Cowan:2010js} and assuming that only
the corresponding FCNC coupling contributes.
The observed and expected \SI{95}{\%} confidence-level (CL) limits on
the branching ratios are shown in \cref{tab:results}, where the
limits on the relevant Wilson coefficients are also reported.
The expected limits on the branching ratios calculated without
systematic uncertainties are lower by \SI{20}{\%} and \SI{25}{\%}
for the \tZu and \tZc couplings, respectively.
The leading systematic uncertainties include the uncertainty
in the SM \SMtZq background normalization and the diboson modeling
uncertainties.
 
\Cref{tab:results} also shows limits on the FCNC \tZu LH and RH
couplings obtained when considering only one SR, either SR1 or SR2, and all
CRs in the likelihood. The results show that SR2, targeting the FCNC
single-top-production signal, contributes more strongly than SR1 to the
combined limits. Separate results for the FCNC \tZc coupling are not
shown, since the limits are dominated by the FCNC-\ttbar-in-decay
signal.
 
\begin{table}[htbp]
\caption{Observed and expected \SI{95}{\%} CL limits on the FCNC
\ttoZq branching ratios and the effective coupling strengths for
different vertices and couplings (top eight rows).
For the latter, the energy scale is assumed to be
$\Lambda_\text{NP} = \SI{1}{\TeV}$.
The bottom rows show, for the case of the FCNC \ttoZu branching
ratio, the observed and expected \SI{95}{\%} CL limits when only
one of the two SRs, either SR1 or SR2, and all CRs are included in
the likelihood.
}
\label{tab:results}
\centering
\begin{tabular}{lcccc}
\toprule
Observable & Vertex & Coupling & {Observed} & {Expected} \\
\midrule
& SRs+CRs & & & \\
\midrule
$\mathcal{B}$(\ttoZq) & \tZu & LH & 6.2$\times 10^{-5}$ & 4.9~$^{+2.1}_{-1.4} \times 10^{-5}$\\
$\mathcal{B}$(\ttoZq) & \tZu & RH & 6.6$\times 10^{-5}$ & 5.1~$^{+2.1}_{-1.4} \times 10^{-5}$\\
$\mathcal{B}$(\ttoZq) & \tZc & LH & 13$\times 10^{-5}$ & 11~{$^{+5}_{-3} \times 10^{-5}$}\\
$\mathcal{B}$(\ttoZq) & \tZc & RH & 12$\times 10^{-5}$ & 10~{$^{+4}_{-3} \times 10^{-5}$}\\
\midrule
$|C_{uW}^{(13)*}|$ and $|C_{uB}^{(13)*}|$ & \tZu & LH & 0.15 & 0.13~$^{+0.03}_{-0.02}$ \\
$|C_{uW}^{(31)}|$ and $|C_{uB}^{(31)}|$   & \tZu & RH & 0.16 & 0.14~$^{+0.03}_{-0.02}$ \\
$|C_{uW}^{(23)*}|$ and $|C_{uB}^{(23)*}|$ & \tZc & LH & 0.22 & 0.20~$^{+0.04}_{-0.03}$ \\
$|C_{uW}^{(32)}|$ and $|C_{uB}^{(32)}|$   & \tZc & RH & 0.21 & 0.19~$^{+0.04}_{-0.03}$ \\
\midrule
& SR1+CRs & & & \\
\midrule
$\mathcal{B}$(\ttoZq) & \tZu & LH & 9.7$\times 10^{-5}$ & 8.6~$^{+3.6}_{-2.4} \times 10^{-5}$\\
$\mathcal{B}$(\ttoZq) & \tZu & RH & 9.5$\times 10^{-5}$ & 8.2~$^{+3.4}_{-2.3} \times 10^{-5}$\\
\midrule
& SR2+CRs & & & \\
\midrule
$\mathcal{B}$(\ttoZq) & \tZu & LH & 7.8$\times 10^{-5}$ & 6.1~$^{+2.7}_{-1.7} \times 10^{-5}$\\
$\mathcal{B}$(\ttoZq) & \tZu & RH & 9.0$\times 10^{-5}$ & 6.6~$^{+2.9}_{-1.8} \times 10^{-5}$\\
\bottomrule
\end{tabular}
\end{table}


\FloatBarrier

 
\section{Conclusions}
\label{sec:conclusions}
 
A search for FCNC processes involving a top quark, an up-type quark
and a \PZ boson is presented.
FCNC \tZq couplings are searched for both in \ttbar decay
events, where one top quark decays according to the SM and the other
one decays as \ttoZq,
and in single top-quark production through the \qtotZ FCNC process,
followed by SM top-quark decay.
The analysis uses \lumi of \pp collision data collected by
the ATLAS experiment at the LHC between 2015 and 2018 at a
center-of-mass energy of \SI{13}{\TeV}.
Events with three leptons, a \Pqb-tagged jet, possible additional jets
and missing transverse momentum are selected.
Multivariate discriminants are used to distinguish signal events from
background events.
 
The data are in good agreement with the SM expectations, and
no evidence of a signal is found.
Limits at \SI{95}{\%} CL are placed on the \ttoZq branching ratios for
both the \tZu and \tZc vertices and for both the RH and LH couplings.
Assuming a LH coupling, the observed limits on the branching ratios
are \SI{6.2e-5}{} for \ttoZu and \SI{13e-5}{} for \ttoZc.
These results for \ttoZu (\ttoZc) improve on the previous observed limits from ATLAS by a factor of 3 (2), and on the previous expected limits by a factor of 5 (3).
These are the most stringent limits to
date. The improvement relative to the previous results comes from
the inclusion of the FCNC-in-single-top-quark-production signal and the usage
of a multivariate analysis in addition to the higher integrated
luminosity.
These results also constrain the values of Wilson coefficients for effective field theory
operators contributing to the \ttoZu and \ttoZc FCNC decays of the top
quark.


\section*{Acknowledgments}


We thank CERN for the very successful operation of the LHC, as well as the
support staff from our institutions without whom ATLAS could not be
operated efficiently.
 
We acknowledge the support of
ANPCyT, Argentina;
YerPhI, Armenia;
ARC, Australia;
BMWFW and FWF, Austria;
ANAS, Azerbaijan;
CNPq and FAPESP, Brazil;
NSERC, NRC and CFI, Canada;
CERN;
ANID, Chile;
CAS, MOST and NSFC, China;
Minciencias, Colombia;
MEYS CR, Czech Republic;
DNRF and DNSRC, Denmark;
IN2P3-CNRS and CEA-DRF/IRFU, France;
SRNSFG, Georgia;
BMBF, HGF and MPG, Germany;
GSRI, Greece;
RGC and Hong Kong SAR, China;
ISF and Benoziyo Center, Israel;
INFN, Italy;
MEXT and JSPS, Japan;
CNRST, Morocco;
NWO, Netherlands;
RCN, Norway;
MEiN, Poland;
FCT, Portugal;
MNE/IFA, Romania;
MESTD, Serbia;
MSSR, Slovakia;
ARRS and MIZ\v{S}, Slovenia;
DSI/NRF, South Africa;
MICINN, Spain;
SRC and Wallenberg Foundation, Sweden;
SERI, SNSF and Cantons of Bern and Geneva, Switzerland;
MOST, Taiwan;
TENMAK, T\"urkiye;
STFC, United Kingdom;
DOE and NSF, United States of America.
In addition, individual groups and members have received support from
BCKDF, CANARIE, Compute Canada and CRC, Canada;
PRIMUS 21/SCI/017 and UNCE SCI/013, Czech Republic;
COST, ERC, ERDF, Horizon 2020 and Marie Sk{\l}odowska-Curie Actions, European Union;
Investissements d'Avenir Labex, Investissements d'Avenir Idex and ANR, France;
DFG and AvH Foundation, Germany;
Herakleitos, Thales and Aristeia programmes co-financed by EU-ESF and the Greek NSRF, Greece;
BSF-NSF and MINERVA, Israel;
Norwegian Financial Mechanism 2014-2021, Norway;
NCN and NAWA, Poland;
La Caixa Banking Foundation, CERCA Programme Generalitat de Catalunya and PROMETEO and GenT Programmes Generalitat Valenciana, Spain;
G\"{o}ran Gustafssons Stiftelse, Sweden;
The Royal Society and Leverhulme Trust, United Kingdom.
 
The crucial computing support from all WLCG partners is acknowledged gratefully, in particular from CERN, the ATLAS Tier-1 facilities at TRIUMF (Canada), NDGF (Denmark, Norway, Sweden), CC-IN2P3 (France), KIT/GridKA (Germany), INFN-CNAF (Italy), NL-T1 (Netherlands), PIC (Spain), ASGC (Taiwan), RAL (UK) and BNL (USA), the Tier-2 facilities worldwide and large non-WLCG resource providers. Major contributors of computing resources are listed in Ref.~\cite{ATL-SOFT-PUB-2021-003}.


\printbibliography

\clearpage
 
\begin{flushleft}
\hypersetup{urlcolor=black}
{\Large The ATLAS Collaboration}

\bigskip

\AtlasOrcid[0000-0002-6665-4934]{G.~Aad}$^\textrm{\scriptsize 100}$,
\AtlasOrcid[0000-0002-5888-2734]{B.~Abbott}$^\textrm{\scriptsize 118}$,
\AtlasOrcid[0000-0002-7248-3203]{D.C.~Abbott}$^\textrm{\scriptsize 101}$,
\AtlasOrcid[0000-0002-2788-3822]{A.~Abed~Abud}$^\textrm{\scriptsize 35}$,
\AtlasOrcid[0000-0002-1002-1652]{K.~Abeling}$^\textrm{\scriptsize 54}$,
\AtlasOrcid[0000-0002-2987-4006]{D.K.~Abhayasinghe}$^\textrm{\scriptsize 93}$,
\AtlasOrcid[0000-0002-8496-9294]{S.H.~Abidi}$^\textrm{\scriptsize 28}$,
\AtlasOrcid[0000-0002-9987-2292]{A.~Aboulhorma}$^\textrm{\scriptsize 34e}$,
\AtlasOrcid[0000-0001-5329-6640]{H.~Abramowicz}$^\textrm{\scriptsize 149}$,
\AtlasOrcid[0000-0002-1599-2896]{H.~Abreu}$^\textrm{\scriptsize 148}$
% ...
\bigskip
\\

% ...
$^{28}$Physics Department, Brookhaven National Laboratory, Upton NY; United States of America.\\
% ...
$^{34}$$^{(a)}$Facult\'e des Sciences Ain Chock, R\'eseau Universitaire de Physique des Hautes Energies - Universit\'e Hassan II, Casablanca;$^{(b)}$Facult\'{e} des Sciences, Universit\'{e} Ibn-Tofail, K\'{e}nitra;$^{(c)}$Facult\'e des Sciences Semlalia, Universit\'e Cadi Ayyad, LPHEA-Marrakech;$^{(d)}$LPMR, Facult\'e des Sciences, Universit\'e Mohamed Premier, Oujda;$^{(e)}$Facult\'e des sciences, Universit\'e Mohammed V, Rabat;$^{(f)}$Institute of Applied Physics, Mohammed VI Polytechnic University, Ben Guerir; Morocco.\\
$^{35}$CERN, Geneva; Switzerland.\\
% ...
$^{54}$II. Physikalisches Institut, Georg-August-Universit\"{a}t G\"ottingen, G\"ottingen; Germany.\\
% ...
$^{93}$Department of Physics, Royal Holloway University of London, Egham; United Kingdom.\\
% ...
$^{100}$CPPM, Aix-Marseille Universit\'e, CNRS/IN2P3, Marseille; France.\\
$^{101}$Department of Physics, University of Massachusetts, Amherst MA; United States of America.\\
% ...
$^{118}$Homer L. Dodge Department of Physics and Astronomy, University of Oklahoma, Norman OK; United States of America.\\
% ...
$^{148}$Department of Physics, Technion, Israel Institute of Technology, Haifa; Israel.\\
$^{149}$Raymond and Beverly Sackler School of Physics and Astronomy, Tel Aviv University, Tel Aviv; Israel.\\
% ...
$^{a}$ Also Affiliated with an institute covered by a cooperation agreement with CERN.\\
$^{b}$ Also at Borough of Manhattan Community College, City University of New York, New York NY; United States of America.\\
$^{c}$ Also at Bruno Kessler Foundation, Trento; Italy.\\
$^{d}$ Also at Center for High Energy Physics, Peking University; China.\\
$^{e}$ Also at Centro Studi e Ricerche Enrico Fermi; Italy.\\
$^{f}$ Also at CERN, Geneva; Switzerland.\\
% ...

\end{flushleft}


\end{document}